\documentclass[useAMS,usenatbib]{mnras}
\usepackage{appendix}
\usepackage{graphicx,dblfloatfix}
\usepackage{amsmath} 
\usepackage{caption}
\usepackage{subcaption}
\usepackage{float}
\restylefloat{figure}
\usepackage{natbib}
\usepackage{booktabs}

\newcommand{\mh}{$ M_{\rm halo}$}
\newcommand{\ms}{$M_{\rm star}$}
\newcommand{\zf}{$ z_{form}$}
\usepackage{amssymb}

\title[Constraining the evolution of the most massive galaxies]{Setting firmer constraints on the evolution of the most massive, central  galaxies from their local abundances and ages}
\author[Stewart Buchan and Francesco Shankar]{Stewart Buchan$^{1}$\thanks{E-mail:
s.w.buchan@soton.ac.uk (SB); f.shankar@soton.ac.uk (FS)} and Francesco Shankar$^{1}$\footnotemark[1]\\
$^{1}$School of Physics and Astronomy, University of Southampton, Southampton, SO17 1BJ, UK}
\begin{document}


\pagerange{\pageref{firstpage}--\pageref{lastpage}} \pubyear{2016}

\maketitle

\label{firstpage}

\begin{abstract}

There is still much debate surrounding how the most massive, central galaxies in the local universe have assembled their stellar mass, especially the relative roles of in-situ growth versus later accretion via mergers. In this paper, we set firmer constraints on the evolutionary pathways of the most massive central galaxies by making use of empirical estimates on their abundances and stellar ages. The most recent abundance matching and direct measurements strongly favour that a substantial fraction of massive galaxies with \ms \textgreater $3 \times 10^{11}\ M_\odot$ reside at the centre of clusters with mass \mh \textgreater $3 \times 10^{13}\ M_\odot$. Spectral analysis supports ages \textgreater 10 Gyrs, corresponding to a formation redshift \zf \textgreater $2$. We combine these two pieces of observationally-based evidence with the mass accretion history of their host dark matter haloes. We find that in these massive haloes, the stellar mass locked up in the central galaxy is comparable to, if not greater than, the total baryonic mass at \zf. These findings indicate that either only a relatively minor fraction of their present-day stellar mass was formed in-situ at \zf, or that these massive, central galaxies form in the extreme scenario where almost all of the baryons in the progenitor halo are converted into stars. Interestingly, the latter scenario would not allow for any substantial size growth since the galaxy's formation epoch either via mergers or expansion. We show our results hold irrespective of systematic uncertainties in stellar mass, abundances, galaxy merger rates, stellar initial mass function, star formation rate and dark matter accretion histories.

\end{abstract}


\section{Introduction}

The mechanisms both to form and evolve massive early-type galaxies are still highly debated. Early-type galaxies (ETG) dominate the red population of objects in the observed bimodal distribution of galaxy colours \citep{Baldry2004,Cassata2008}. However, how ETGs form, evolve and transition in the colour-mass plane remains unclear \citep[][and references therein]{Woo2015}.

Significant effort has been put into probing the evolution of the most massive galaxies both observationally and theoretically. From the observational side, a renewed interest in this field emerged with the advent of large and deep galaxy surveys such as SDSS \citep{Ahn2014}, COSMOS \citep{Scoville2007}, BOSS \citep{BOSS2013} and CANDELS \citep{Grogin2011}. A key result of these surveys is the observed evolution of the size-mass relation of ETGs. Massive, early type galaxies are observed to be progressively more compact at higher redshift as compared to galaxies with the same mass in the local universe \citep{VanDokkum2010,Huertas-Company2013}. This size evolution could be triggered by later mergers \citep[e.g.,][]{Naab2009,Shankar2013}, gas accretion \citep{Dekel2009} and/or nearly adiabatic expansion due to quasar mode feedback and/or stellar winds \citep{Fan2008,Damjanov2009}. However, despite significant progress, issues such as ``progenitor bias'' \citep[e.g.,][]{VanDokkum1996, Saglia2010,Carollo2013,Shankar2015}, the role of environment \citep{Poggianti2006,Shankar2013,Delaye2014,Shankar2014a,Stringer2015} and observational systematics such as cosmic variance and stellar mass estimates \citep{Marchesini2009,Bernardi2013}, hinder a clear interpretation of the observed mass and size evolution, especially at the extreme high-mass end of the stellar mass function \cite[e.g.,][]{Marchesini2014, Shankar2014a,Leauthaud2016,Bernardi2016}.

From the theoretical side, evolutionary models have not converged on one clear picture of massive galaxy evolution. Semi-analytic models are an effective way of probing the diverse physical processes that are believed to drive galaxy formation and evolution \citep{Cole2000,Baugh2006,Guo2011,Benson2012,Lacey2015}. These models, sometimes based on very different input assumptions, can offer degenerate solutions in reproducing a handful of key statistical properties such as the galaxy stellar mass function \citep[see review discussion in][]{Mo2010}. Broadly speaking, semi-analytic models predict two conflicting evolutionary pathways, one where dry mergers dominate the evolution \citep{DeLucia2011,Gonzalez2011,Guo2011,Shankar2013,Wilman2013} and one where in-situ processes are more important \citep{Lapi2011,Ragone-Figueroa2011,Chiosi2012,Merlin2012,Posti2014}.

Hierarchical merger models predict that massive galaxies have assembled most of their final stellar mass via a sequence of mergers following their host dark matter haloes \cite[e.g.,][]{Naab2009,Shankar2009,VanDokkum2010,Guo2011,Shankar2013,Montes2014}. Indeed massive galaxies must have merged at some point as tidal tails and concentric shells are observed around massive, local galaxies \citep{Duc2015}. The rate at which galaxies merge is usually observationally inferred by looking for pairs of galaxies in close spatial proximity \cite[e.g.,][]{Hopkins2010}. However, this rate is non-trivial to quantify \citep[e.g.,][]{Conselice2014} as a number of systematics may effect the result, from the assumptions on dynamical friction timescales, to the determination of spectroscopic pairs. Consequently, the true role of mergers in shaping massive galaxies remains still uncertain.

In-situ galaxy evolutionary models claim instead that massive galaxies formed and assembled most of their final stellar mass in strong bursts of star formation at high redshifts. These starbursts can have star formation rates as high as several thousands of solar masses per year \citep{Chapman2005}. After the starburst has quenched, possibly induced by an efficient quasar mode feedback, the galaxy is assumed to evolve almost passively until the present day \citep{Granato2004,Granato2006,Carollo2013,Zolotov2015}. ETGs are observed to be enhanced in alpha-elements relative to their iron content which is evidence for short bursts of intense star formation \citep{Thomas2005,Pipino2009,Conroy2014,Citro2016}.

Hydrodynamical zoom simulations \citep[e.g.,][]{Hirschmann2012} have converged on the idea that there are two phases to massive galaxy evolution where in-situ star formation dominates the early assembly and mergers become more important at lower redshifts \citep{Naab2009,Oser2010}. Hydrodynamical simulations in a full cosmological box continue to support this two-stage evolutionary patten at least for the most massive galaxies \citep{Hirschmann2012,Torrey2015,Welker2015}. However, the relative roles of in-situ versus late assembly remains poorly constrained observationally.

In recent years, a number of semi-empirical approaches have been put forward to more securely probe and constrain the possible evolutionary pathways of massive galaxies. For example, \cite{VanDokkum2010}, \cite{Marchesini2014} and \cite{Huertas-Company2015} have adopted number conservation techniques to track the putative main progenitors of massive galaxies. Other techniques are based on continuity equation models for the stellar population \citep[e.g.,][]{Peng2010,Aversa2015}. Also, \cite{Lidman2012} and \cite{Shankar2015} followed the main progenitor track of the host haloes to identify potential proto-galaxies as progenitors. All of these semi-empirical approaches broadly agree in assessing the primary role of in-situ growth for galaxies below \ms$\lesssim 10^{11}\ M_\odot$. However, models become generally more discordant when predicting the evolution of the most massive galaxies. One of the main reasons for such discrepancies can be traced back to the growing significance of the systematics associated with observations such as surface brightness variations, estimates of the proper background, cosmic variance, stellar mass estimates, the number of mergers and the initial mass function \citep[][]{VanDokkum2010,Marchesini2009,Behroozi2013,Maraston2013,Bernardi2014,Shankar2014a,Aversa2015,Leauthaud2016,Bernardi2016b}. In particular, \cite{Bernardi2016} have recently shown that even when homogeneous measurements are carried out at different redshifts, a clear understanding of the evolution of the most massive galaxies still remains elusive. 

The aim of this paper is to set more stringent and secure constraint on the evolution of the most massive, central galaxies in the local universe for \ms \textgreater $3 \times 10^{11}\ M_\odot$ for which data are still incomplete and/or uncertain, especially at high redshifts. In this work, we use a series of observationally-driven models that, by design, rely on very few assumptions and thus provide us with constraints less clouded by more complex modelling.

This paper is structured as follows: In section 2 we give an overview of our methodology and describe our sample selection. In section 3 we discuss the constraints we set on the assembly scenario of massive galaxies. In section 4 we investigate the relative importance of in-situ processes and mergers in driving the evolution of massive ETGs in a late assembly scenario using both observationally informed models as well as a full cosmological, semi-empirical models. 

We adopt a flat $\Lambda CDM$ cosmological with $\Omega_M = 0.3$, $h = 0.7$, $\Omega_B = 0.045$, $\sigma_8 = 0.8$, $d_c^0 = 1.69$, and assume a Chabrier initial mass function \citep[IMF:][]{Chabrier2003}. Throughout this paper, we define the halo mass as \mh$=M_{200c}$, 200 times the critical density at redshift $z$.

\section{Overview and Methodology}

In a dark-matter dominated universe, large scale structures are formed from the collapse of primordial density fluctuations \citep{White1991}. Over cosmic time, cold primordial gas condenses within these density perturbations forming the baryonic portion of galaxies \citep[see][for a detailed review]{Mo2010}. However, the processes of turning primordial gas into the galaxies we observe today are still debated. Here, we circumvent the complexities of baryon physics by tracing the evolution of the host dark matter haloes, which is more transparent and secure as it relies only on gravitational physics. We then map galaxies to haloes in a statistical sense using semi-empirical relationships. We selected dark matter haloes from the dark matter-only Bolshoi simulation \citep{Klypin2011}, which provides the full dark matter merging history.

\begin{figure}[h!]
\centering
\includegraphics[width=0.5\textwidth]{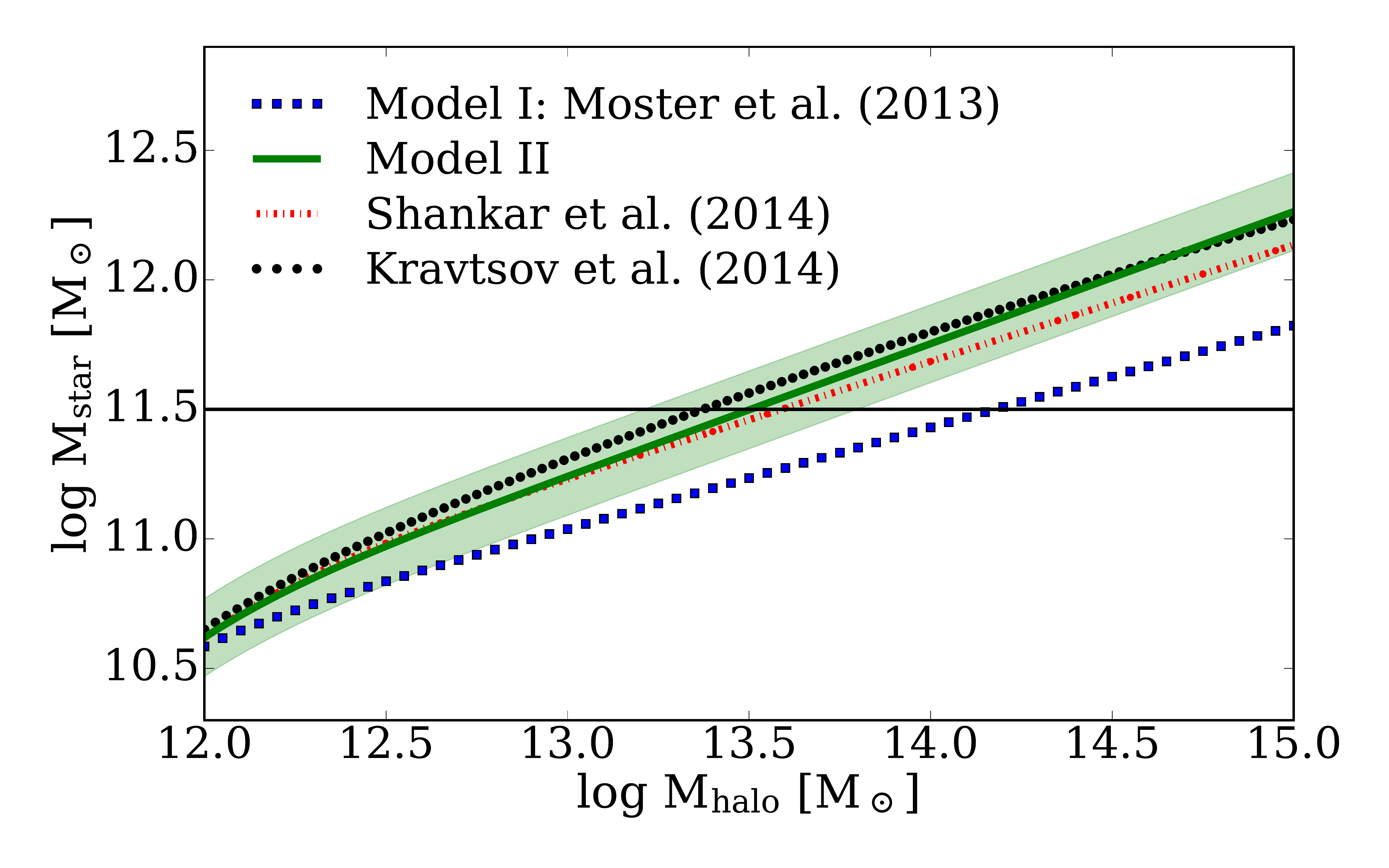}
\caption{A comparison among estimates of the stellar mass-to-halo mass relations. In this work, we adopt the relations by \protect\cite{Moster2013} as well a steeper version which matches the relations by \protect\cite{Shankar2014b} and \protect\cite{Kravtsov2014}.}
\end{figure}

More specifically, to set more stringent constraints on the evolutionary patterns of local massive galaxies, we use an observationally-driven model that works as follows:

\begin{enumerate}
\item  We extract all central haloes from the Bolshoi simulation and assign them a stellar mass, \ms , using the latest renditions of the \ms\ - \mh\ relation.
\item  We select those haloes hosting central galaxies with $\log\ ($\ms $)>11.5\ M_\odot$. 
\item  For each halo, we track its progenitors backwards in time until the putative formation epoch, \zf .
\item  We estimate their total baryonic mass and stellar mass at \zf\ from the global baryon fraction and from abundance matching relations, respectively.
\item  We finally compare the estimated baryonic mass to their descendent galaxy's stellar mass at $z=0$.
\end{enumerate}

\subsection{Selecting descendant galaxies at z=0}

Techniques such as abundance matching allow us to connect galaxies to their dark matter haloes in a statistical sense. Abundance matching works by matching the cumulative number densities of dark matter haloes to the observed number densities of galaxies. This is a powerful technique in predicting the mean stellar content of dark matter haloes, especially for massive, central galaxies with \ms$>2\times 10^{11} M_\odot$, where the scatter in stellar mass reduces to $\leq0.15dex$, and the dispersion in assembly histories due to, e.g., environment, age spread, specific star formation history, becomes less important \citep{Shankar2014a, Gu2016, Clauwens2016}. In this work, we use the parameterizations of the stellar mass to halo mass relation of \cite{Moster2013}:

\begin{equation}
\frac{M_{\rm star}}{M_{\rm halo}}=2N\left[\left(\frac{M_{\rm halo}}{M_1}\right)^{-\beta}+\left(\frac{M_{\rm halo}}{M_1}\right)^\gamma\right]^{-1}
\end{equation}

\noindent
where $M_1$, $N$, $\beta$ and $\gamma$ are constants. 

One of the main sources of systematic uncertainties in the high mass end of the \ms\ -\mh\ relation comes about from the exact shape of the stellar mass function. It has recently been shown that the high mass end of the stellar mass function has significantly higher abundances than earlier measurements \citep{Bernardi2013,D'Souz2015,Bernardi2016b,Thanjavur2016}. \citet{Bernardi2016b} in particular, have shown that possible systematics in photometry are now of the order $\sim 0.1dex$. Recent results by \cite{Kravtsov2014} and \cite{Shankar2014b}, based on the new stellar mass function of \cite{Bernardi2013}, coupled with direct measurements of the stellar masses and host halo masses of individual brightest group and cluster galaxies, conclude that the mean stellar mass of massive central galaxies is systematically a factor of $\sim$3-4 higher at fixed halo mass than previously estimated by, e.g., \cite{Behroozi2013} and \cite{Moster2013}. In what follows, to bracket the possible residual systematics in the \ms\ -\mh\ relation, we will adopt Equation 1 with \textit{both} the original parameters found by \citet[][hereafter model I]{Moster2013}  as well as with updated parameters to match the results of \cite{Kravtsov2014} and \cite{Shankar2014b} in the stellar mass range of interest in this paper (hereafter model II; see table 1 for the new parameters).

\begin{table}
\centering
\caption{The parameters used in equation 1, the \ms\ - \mh\ relation for both the original Moster et al. (2013) relation as well as our steeper version.} 

\label{my-label}
\begin{tabular}[c]{@{}cllll@{}}
\toprule
 & $M_1$ & $N$ & $\beta$ & $\gamma$ \\ \midrule
Model I & 11.59 & 0.0351 & 1.376 & 0.608 \\
Model II & 11.70 & 0.0380 & 1.25 & 0.490 \\
\bottomrule
\end{tabular}
\end{table}

Figure 1 shows a comparison between models I, II and the latest relations by \cite{Shankar2014b} and \cite{Kravtsov2014}. We choose to specifically consider galaxies with \ms $>3\times 10^{11}\ M_\odot$ as this is the threshold in stellar mass where there is most disagreement in the assembly history among different galaxy evolutionary models \citep[e.g.,][]{Bernardi2016}. This stellar mass cut of $\log\ ($\ms$)>11.5\ M_\odot$ is shown as a horizontal line in Figure 1.

When applying the \ms\ - \mh\ relation to high redshift progenitor haloes, we keep the original redshift dependence of \cite{Moster2013} for both models I and II. Maintaining the Moster et al. (2013) formalism has the advantage to directly extend the abundance matching to $z>0.5-1$ a redshift regime beyond the one probed by \cite{Kravtsov2014} and \cite{Shankar2014b}. It is important to note that at higher redshift, we are probing haloes with \mh $\sim 10^{12}\ M_\odot$, which sit around the knee of the stellar mass function and are thus significantly less prone to the above mentioned systematic uncertainties (for example, in stellar masses) characterizing the high-mass end of the \ms\ - \mh\ relation. We anyway stress that our main conclusions do not depend on the exact redshift dependencies in \cite{Moster2013}. We also note that \cite{Behroozi2013} are consistent with \cite{Moster2013} within the uncertainties quoted in this paper.

\begin{figure*}[t!]
\centering
\includegraphics[width=.99\textwidth]{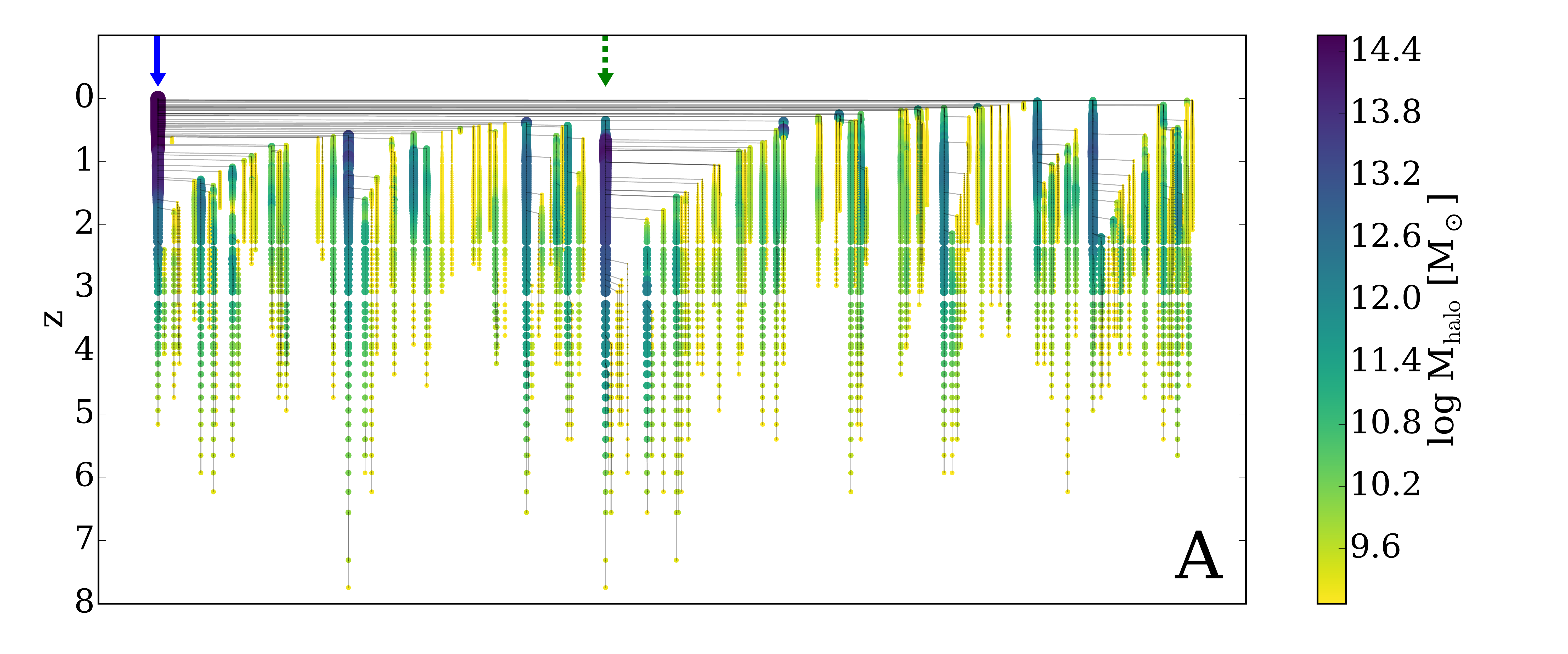}
\includegraphics[width=.99\textwidth]{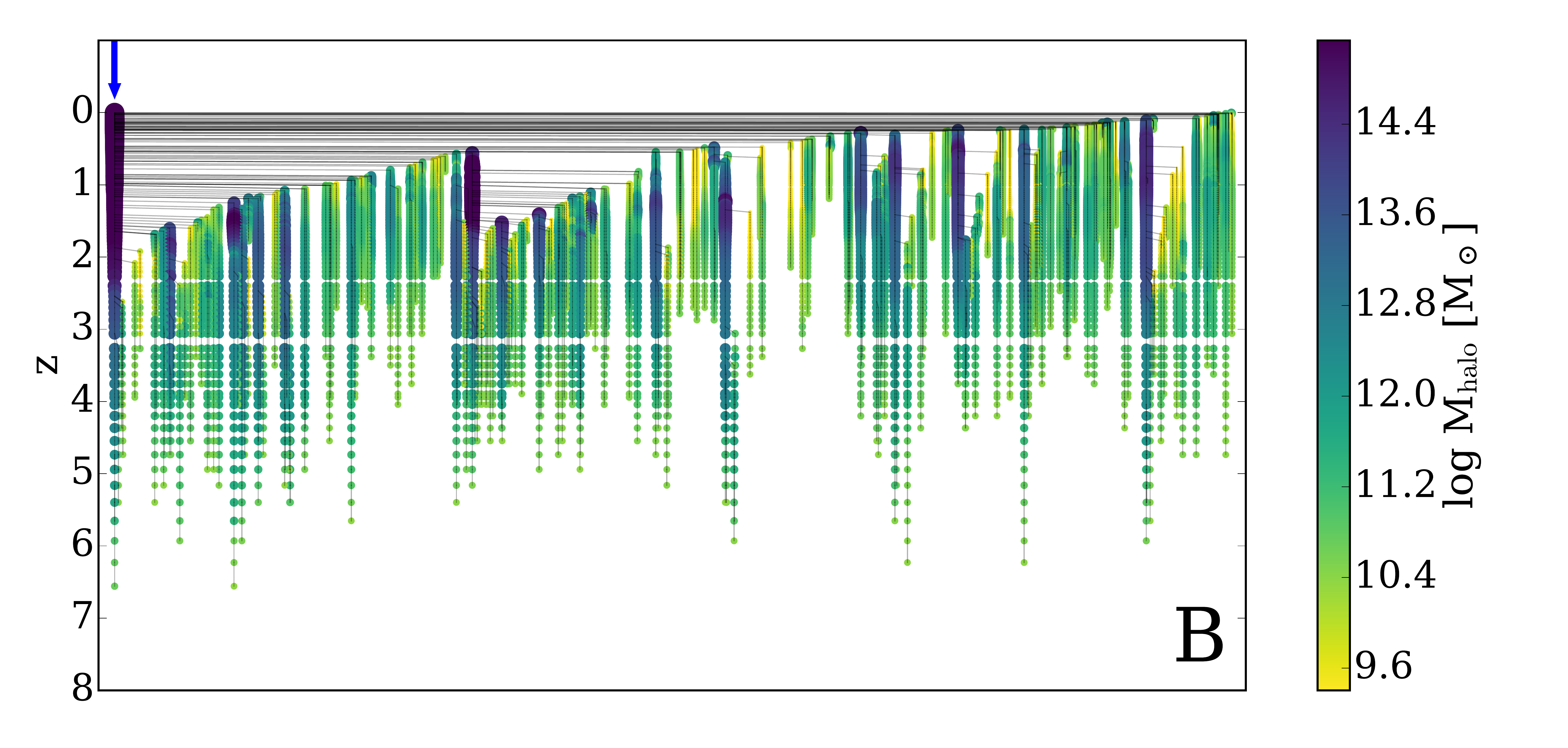}
\includegraphics[width=0.49\textwidth]{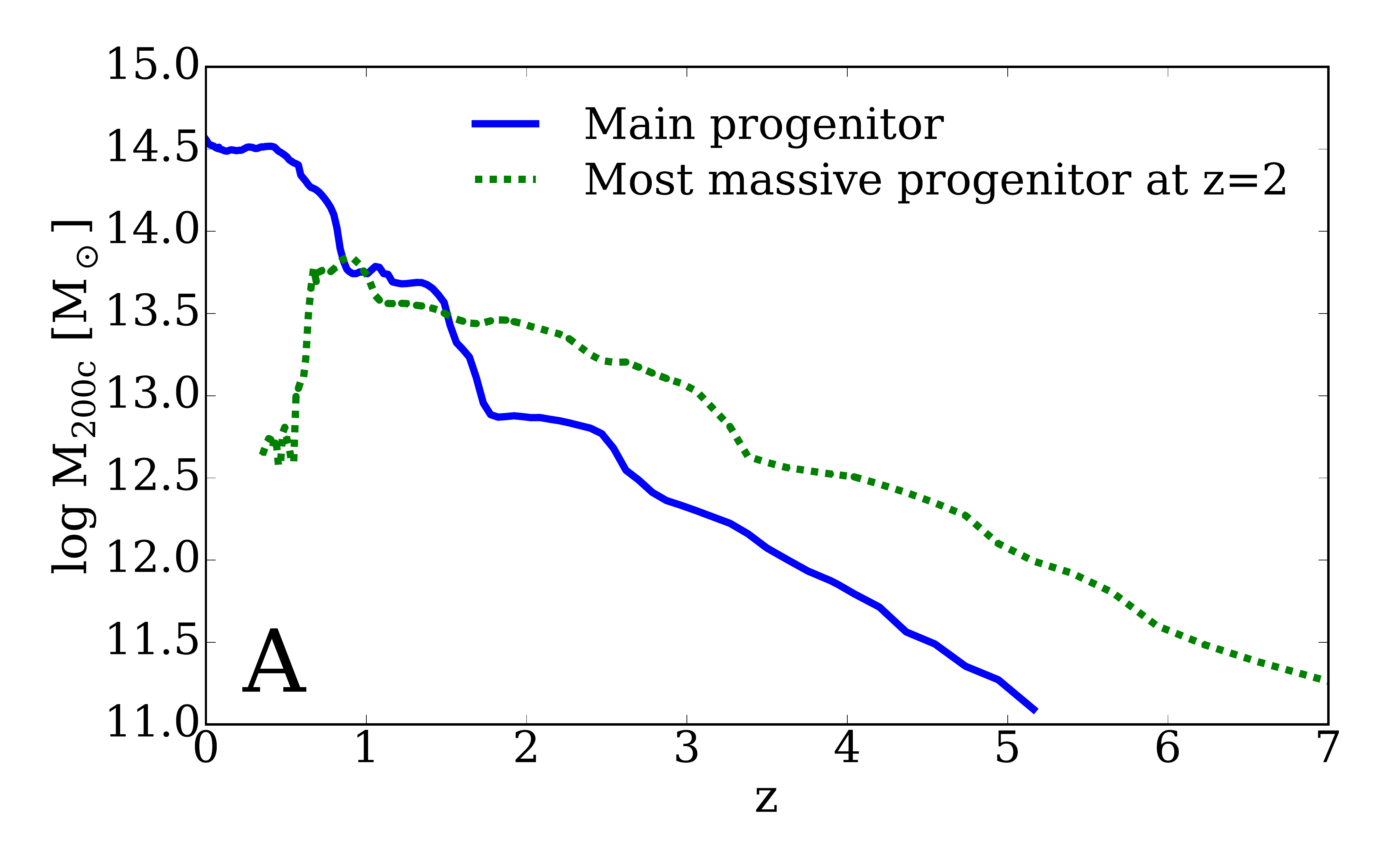}
\includegraphics[width=0.49\textwidth]{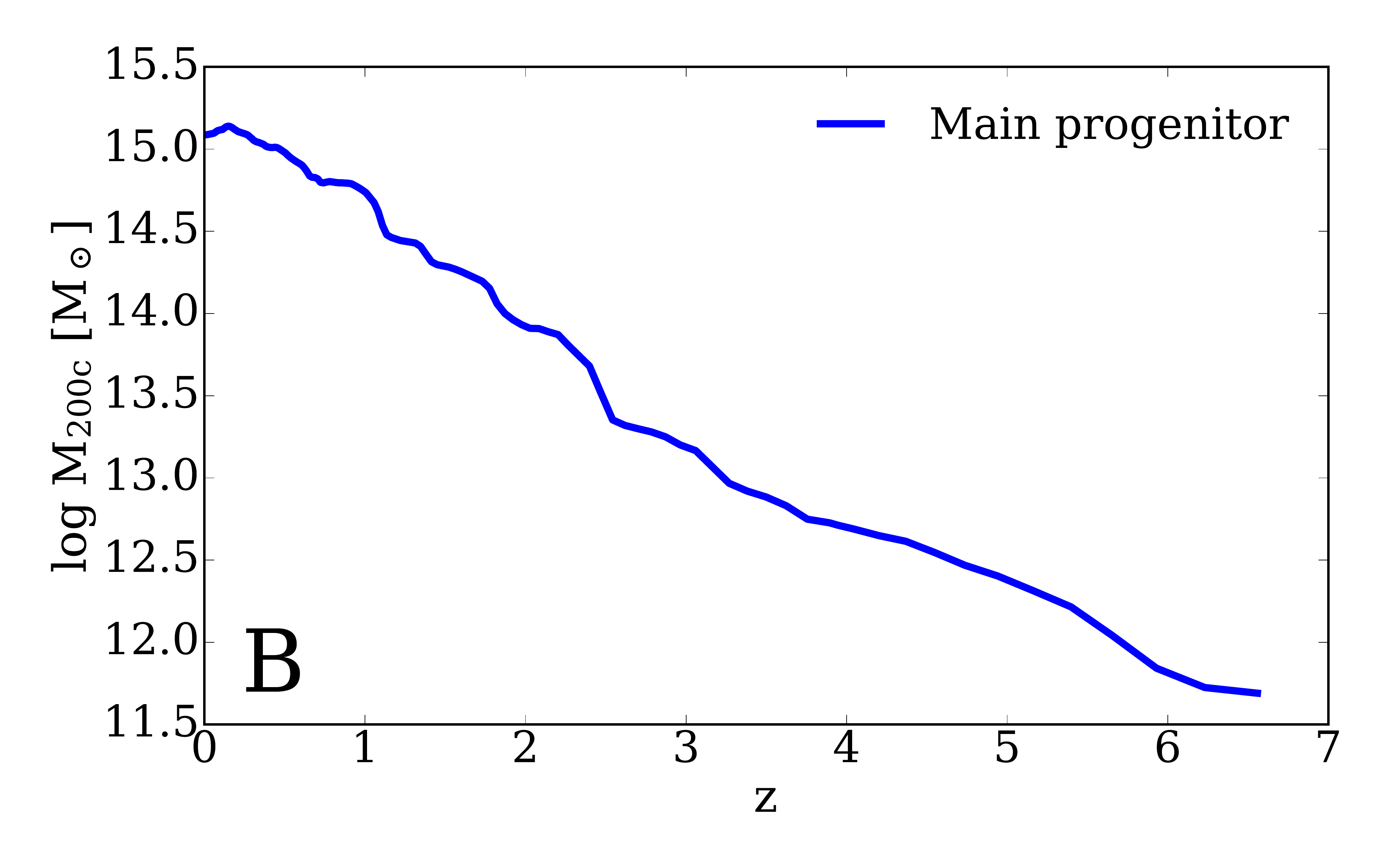}

\caption{The top panels are visualizations of two dark matter merger trees and bottom panels show the mass evolutions of the main progenitor (solid) and the most massive progenitor at $z=2$ (dashed line). The panels labelled A show the evolution of a halo where the main progenitor halo is not the most massive halo at high redshift. The panels labelled B instead shows a more idealised case where the main progenitor is the most massive at all the redshifts we consider.}
\end{figure*}

\subsection{Selecting their progenitors}

The progenitor of a massive galaxy is usually considered to follow the so-called ``main progenitor'' halo. This is defined as the most massive progenitor of a parent halo \citep{Jiang2014}. The main progenitor branch is therefore a chain of haloes constructed by finding the most massive progenitor of the previous main progenitor, starting at $z=0$ and working backwards in time. However, this definition does not necessarily imply that the main progenitor is the most massive progenitor at all times. In fact, studies such as those by \cite{Lapi2013}, which are based on the excursion-set formalism \citep{Bond1991}, have pointed out that possibly the most massive progenitor haloes at \zf\ are more relevant than the main progenitor for the evolution of today's central, massive galaxies. This is an important distinction to make in this paper as a more massive halo has more potential to form a more massive galaxy in a single burst.

To visualise the difference between the main progenitor and most massive progenitor branches, Figure 2 shows the merger trees of two representative dark matter haloes in the Bolshoi simulation. The panels labelled ``A'' show a merger tree whereby the main progenitor is \textit{not} the most massive progenitor at all redshifts. The panels labelled ``B'' instead show a more idealised case where the main progenitor is the most massive progenitor at all epochs. The top two panels show a visualisation of the merger trees with redshift along the $y$-axis and the branches of the tree separated out along the $x$-axis. The main progenitor branch is positioned at the far left-end of the plots and is indicated with a blue, solid arrow. We also indicate the branch which contains the most massive progenitor at $z=2$ with a green, dashed arrow. The horizontal lines show merging events between the branches. The size of the circles is proportional to the mass of the progenitors at that redshift, as encoded in the colour legend. The bottom panels then show the mass evolution of the main progenitor and of the most massive $z=2$ progenitor.

The relevant question we need to answer is how frequent the main progenitor remains indeed the most massive progenitor at all times, as in the panels labelled ``A'' of Figure 2.  To this purpose, we carefully analyse the merger trees of each halo in our sample\footnote{We verified that most massive progenitor distributions are very similar in the MultiDark Planck 2 simulation \citep{Klypin2016} which has a larger volume and different cosmology than Bolshoi.} of galaxies with $\log\ ($\ms $)>11.5\ M_\odot$. In Figure 3, we show the mass functions of the main progenitors (solid) and most massive progenitors (dashed) at the labelled redshifts. From this figure, we conclude that the choice in definition of progenitors has little impact on the mean evolution in halo mass and at most is only relevant for the low mass wings of the distributions at $z>3$. This is because only $\lesssim 25\% $ of the haloes follow the ``most massive'' progenitor track. We checked that this conclusion still holds even if we restrict the analysis to only the most massive haloes in our sample with $\log\ ($\mh$)>14\ M_\odot$ where the effect could be most prominent \citep{Lapi2013}. In the following, we use the main progenitor as our reference, though we also show results using the most massive progenitor, where relevant.

\begin{figure}[h!]
\centering

\includegraphics[width=0.5\textwidth]{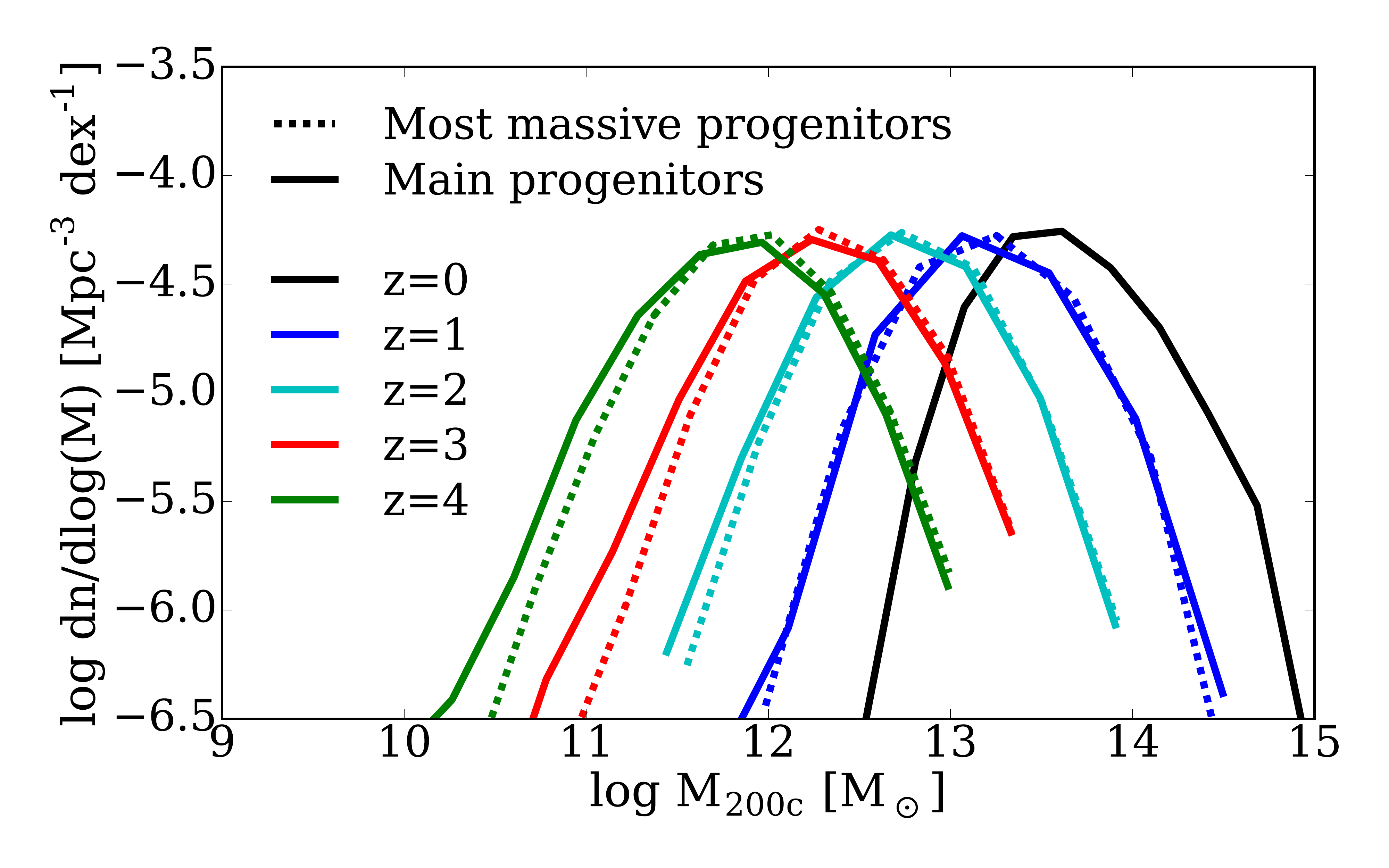}
\caption{The mass functions of the progenitor haloes that contain galaxies with $\log\ ($\ms $)>11.5\ M_\odot$ at $z=0$. The solid and dashed lines show the main progenitor and most massive progenitor mass functions at different redshifts, as labelled.}
\end{figure}

\subsection{Ages of massive, early type galaxies: selecting a formation epoch.}

From stellar population synthesis modelling it is possible to estimate the mass-weighted age of the stars within a galaxy. The general method to constrain the age of a galaxy is to fit the galaxy's spectra with either a single or a combination of synthetic stellar populations with varying star formation rates, for a given initial mass function \citep{Vincoletto2012, McDermid2015, Mendel2015, Citro2016}. For the galaxies of interest to us $($\ms$> 3 \times 10^{11}\ M_\odot)$, the majority of the stars form at or above \zf $=2$, with less massive galaxies having, on average, younger stellar populations \citep{Thomas2005}. This is the so-called `top-down' mass assembly scenario for massive ETGs.

We explore the consequences of a formation redshift, \zf $=2-4$. For reference, \cite{McDermid2015} find that $\geq$ 50\%(90\%) of the stellar mass is older than $z=3(2)$ for the stellar mass we consider in this work\footnote{Note that strictly speaking the stellar masses in \cite{McDermid2015} are dynamical masses from jeans modelling. \cite{Cappellari2013} discuss that these masses are closer to those measured assuming a Salpeter IMF. This would imply that the galaxies we have selected for this work have higher stellar mass at fixed halo mass and hence we could be, if anything, underestimating their age.}. At these formation epochs, we compare the amount of baryons in the progenitor haloes to the stellar content of the descendants.

\section{results}

\subsection{Constraints on the assembly scenario}

\begin{figure}[h!]
\centering
\includegraphics[width=0.5\textwidth]{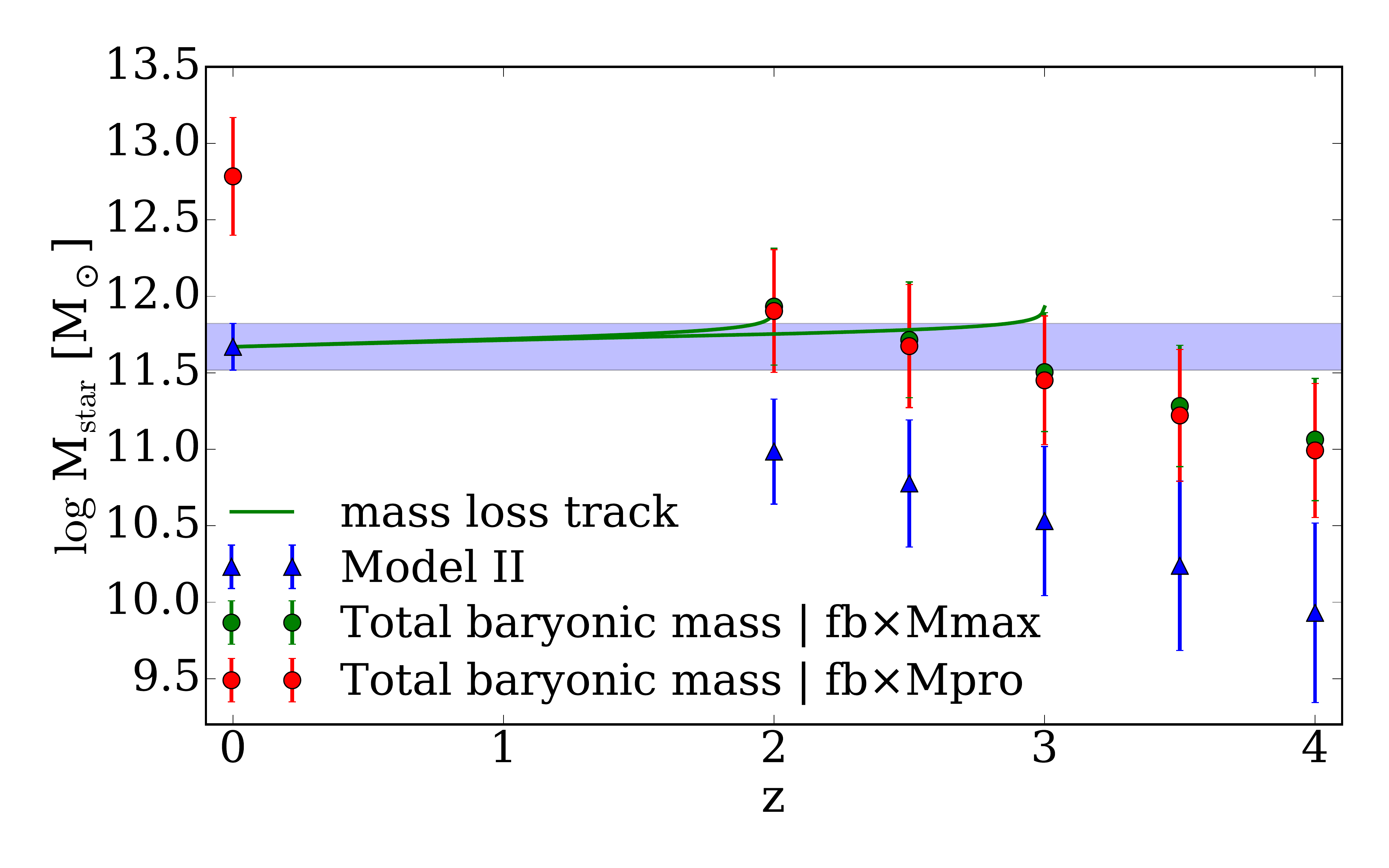}
\includegraphics[width=0.5\textwidth]{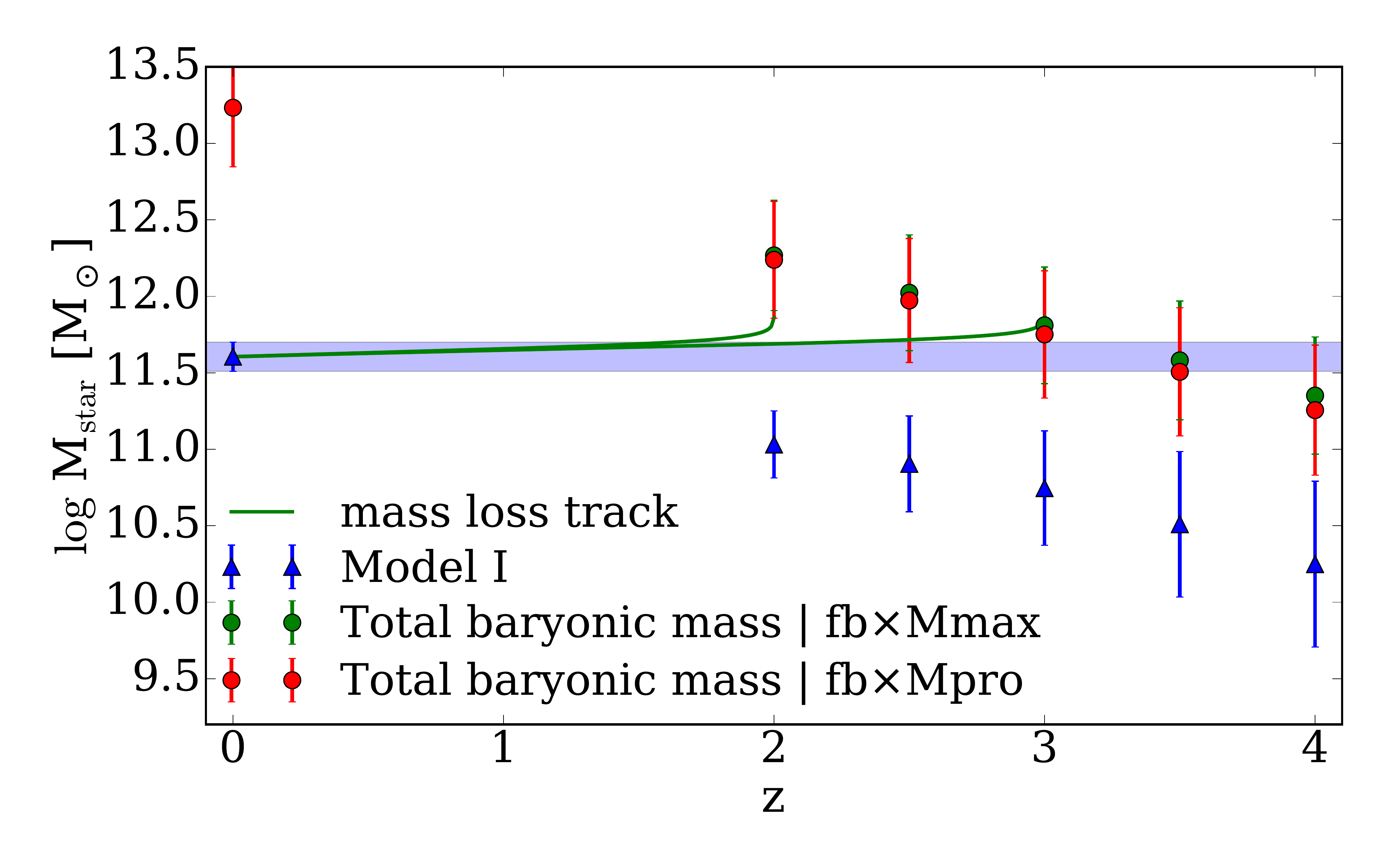}
\caption{A plot showing the predicted mean stellar mass of galaxies with $\log\ ($\ms $)>11.5\ M_\odot$ halo at $z=0$ (band) using the original \protect\cite{Moster2013} \ms\ - \mh\ relation (bottom) and a modified version to match the latest relation from \protect\cite{Shankar2014b} (models I and II, respectively). We show the mean and 1$\sigma$ of the total baryonic mass associated with the the main progenitor haloes (red circles) and most massive progenitors (green circles). For reference, we also show the stellar mass estimates of galaxies in the respective haloes using the given relations (blue triangles).}
\end{figure}

Having constructed our $z=0$ dark matter halo catalogue and traced their progenitors to \zf, we are now in a position to compare the baryonic content between progenitor haloes at \zf\ and descendent haloes at $z=0$. The total baryonic mass is computed using the cluster baryon fraction extracted from numerical simulations. \cite{Crain2007}, in particular, find that the baryon fraction inside the virial radius of dark matter haloes is 90\% of the cosmic mean fraction, independent of halo mass and redshift

\begin{equation}
M_{baryon}=0.9\times M_{halo}\times f_b
\end{equation}

\noindent
where $f_b$ in Eq. 2 is the global mean ratio between the baryon and dark matter density. We show the implied total baryon masses in Figure 4 for the main progenitor haloes (red circles) and most massive progenitor (green circles) at the putative formation epochs \zf $=2$ to $4$ in steps of 0.5. The blue triangles show instead the mean stellar mass computed along the main progenitors adopting model I (the \ms\ - \mh\ relation using the original parameters found by \cite[][botton panel]{Moster2013} and Model II (the updated parameters, top panel). Additionally, for reference we show the mean mass evolutionary tracks expected in a passive evolutionary model, taking into account the ageing stellar population, following Eq. 14 by \cite{Behroozi2013} who also use a Chabier IMF and \cite{Bruzual2003} stellar evolution tracks. The green solid lines show two evolutionary tracks that are needed to reproduce the $z=0$ stellar mass distribution starting at \zf $=2$ and $3$.

The first point to make regarding Figure 4 is that the local stellar mass of massive ETG's is comparable to, if not greater than, the total baryonic mass contained in their main  progenitor haloes at \zf. This is especially evident for the top panels, which utilise the steeper high mass slope in model II to map lower mass haloes at fixed stellar mass (Figure 1), while it only becomes evident at $z>3$ in model I. Thus the updated abundance matching relations would imply that if a galaxy is formed via a strict monolithic collapse at the epoch of formation, star formation should have been extremely efficient, if not 100\%, to account for the (high) stellar mass content observed today in the descendant haloes. Moreover, this scenario would preclude any substantial loss of baryons due to stellar winds and/or quasar mode feedback during this rapid star formation phase. We note that this result is not changed and, in fact, possibly strengthened by assuming a Salpeter IMF, which some authors \citep[e.g.,][]{Cappellari2013} have suggested to be more representative of massive galaxies. A Salpeter IMF would imply a higher stellar mass at $z=0$ for the same set of host haloes of interest here (\mh$>3\times 10^{12}\ M_{\odot}$), worsening the tension with the available baryons at \zf.

\subsection{Are today's central, massive galaxies just outliers at the epoch of formation?}

Figure 4 also shows that $z=0$ stellar masses are a factor of at least five times larger than the typical stellar mass of the progenitor galaxies at \zf\ as predicted by abundance matching (blue triangles). There are two possible conclusions from this finding. If the progenitors of massive galaxies at \zf\ are representative of galaxies at that halo mass, then clearly a later mass growth is needed to match the stellar mass of the descendants to their local counterparts. On the other hand, the progenitors of very massive galaxies might not be representative of the general population of galaxies at \zf\ at fixed halo mass. For instance, they could be extreme outliers with a stellar mass much greater than what is predicted by abundance matching relations. This could arise in  strictly monolithic models where stars are formed in an extremely efficient and fast mode around \zf.

\begin{figure}[h!]
\centering
\includegraphics[width=0.45\textwidth]{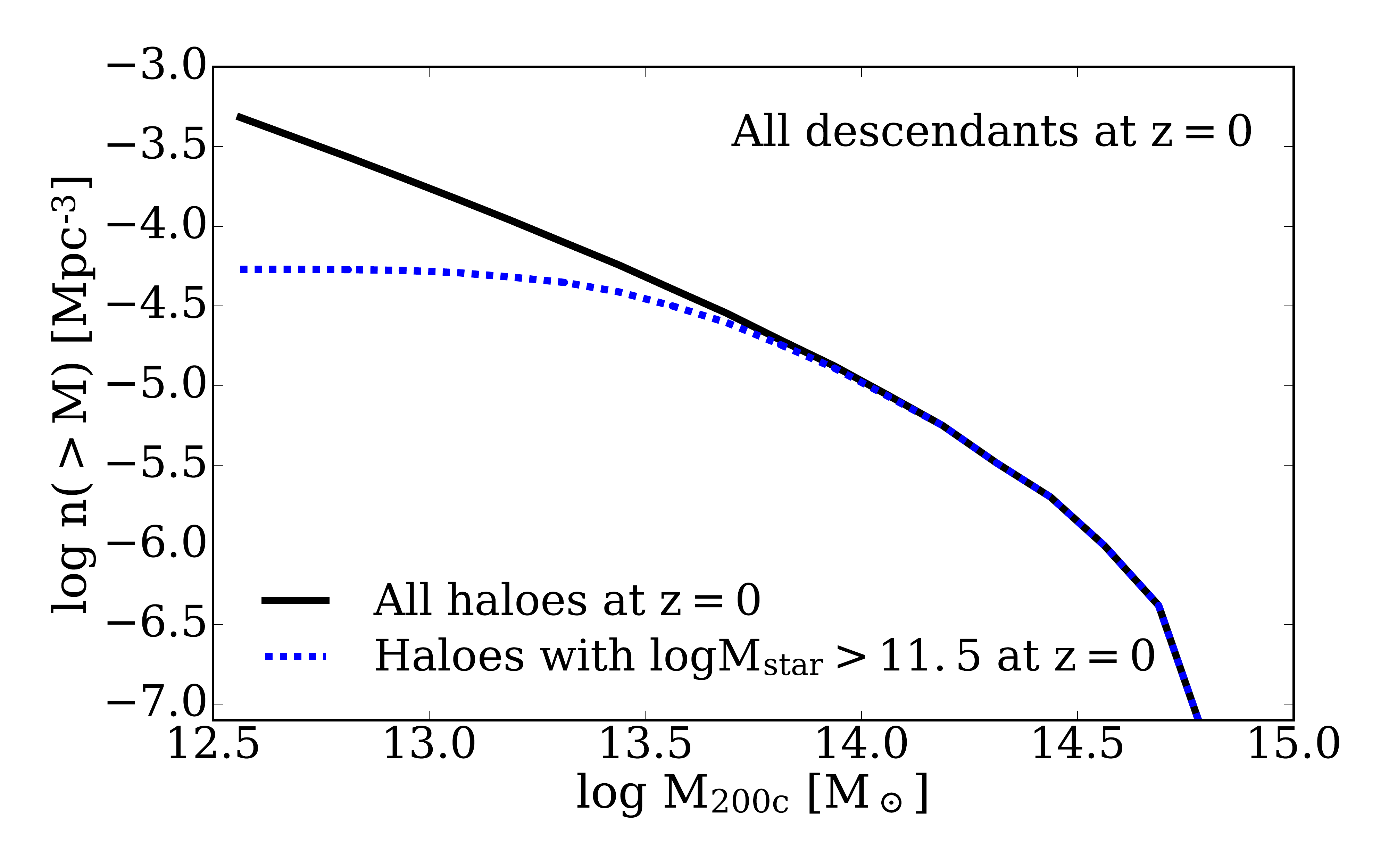}
\includegraphics[width=0.45\textwidth]{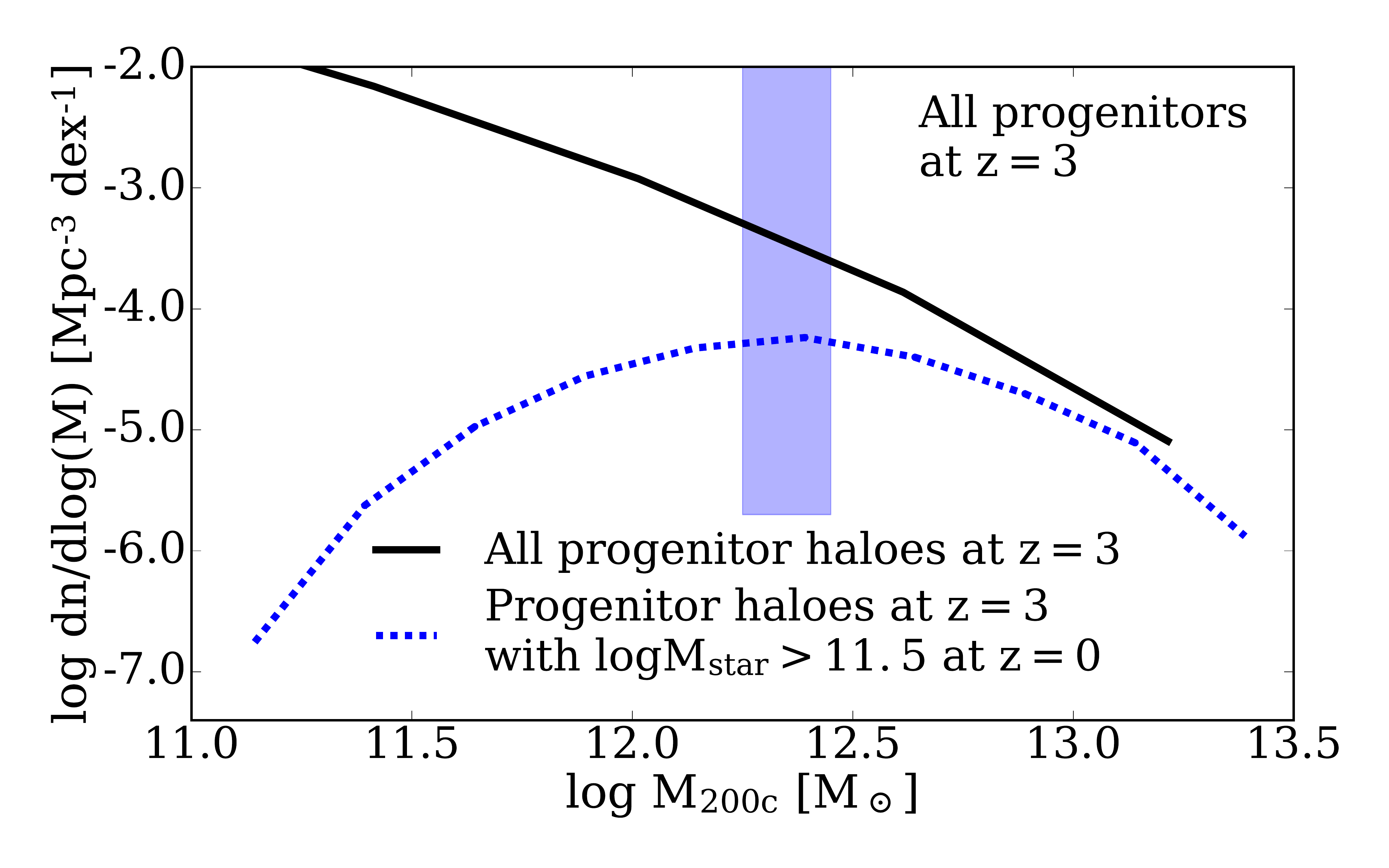}
\includegraphics[width=0.45\textwidth]{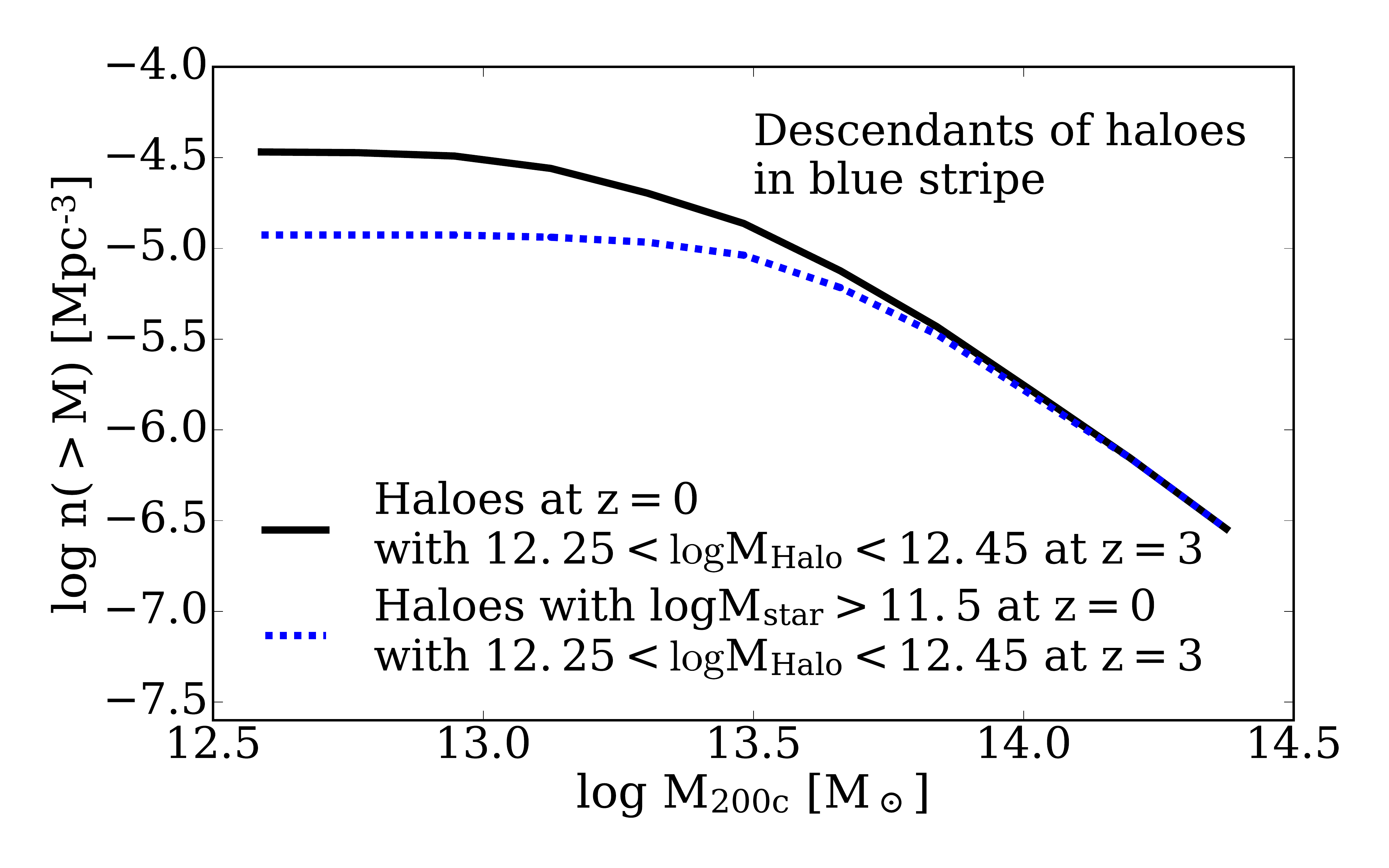}
\caption{Top: the integrated number density of dark matter haloes in the Bolshoi simulation. The black line shows the distribution for all haloes in the simulation and blue shows those which are included in our stellar mass selection. Middle: the differential mass function of dark matter haloes at z=3 which survive to z=0. The black line shows the total halo distribution and blue are those haloes which enter our stellar mass selection. Bottom: we apply a halo mass selection at z=3 (highlighted in the middle panel) and follow those haloes forward to z=0. The black shows the distribution of haloes when selected from the total distribution and blue when selected from those which are included in our stellar mass cut.}
\end{figure}

To probe the latter possibility, we proceed as follows. In the top panel of Figure 5 we show as black solid and blue dashed lines respectively, the full $z=0$ cumulative mass functions of all massive dark matter haloes with \mh$>3\times10^{12}\ M_{\odot}$, and of those hosting our stellar mass-selected sample. We then trace the main progenitors of our selected galaxies back to $z=3$. In the middle panel of Figure 5, we show the differential mass functions of the main progenitors of all massive $z=0$ haloes (solid black) and of those in our stellar mass selection (dashed blue). Here, it is evident that the mean mass of these progenitor haloes at $z=3$ is \mh$\sim 2\times 10^{12}\ M_\odot$, consistent with Figure 3, which we highlight as a blue band in the middle panel. We select those haloes at $z=3$ which are between $12.25<\log\ ($\mh$)<12.45$ and follow them forward to $z=0$. The black solid line in the bottom panel of Figure 5 is the cumulative number density at $z=0$ of haloes which have a $z=3$ progenitor mass \mh$\sim 2\times 10^{12}\ M_\odot$. The blue dashed line instead are haloes which are in the same mass range at z=3 but also become massive enough to enter our $z=0$ stellar mass selection.

By comparing the cumulative mass functions, we find that on average, only $\sim 30\%$ of the haloes with mean host mass \mh$\sim 2\times 10^{12}\ M_\odot$ at $z=3$ will host galaxies with $\log\ ($\ms$)>11.5\ M_{\odot}$ at $z=0$. Our selected brightest cluster galaxies are thus only a relatively minor fraction of the galaxies residing in \mh$\sim 2\times 10^{12}\ M_\odot$ at $z=3$. We thus conclude that the progenitors of massive galaxies could indeed be outliers with respect to the general population of galaxies at \zf$=3$ and with \mh$\sim 2\times 10^{12}\ M_\odot$. However, they represent the \textit{majority} of galaxies that will end up as centrals in haloes with \mh$>5\times 10^{13}\ M_\odot$, within the mass scale of massive groups and clusters.

\section{Discussion}

From the previous sections we can conclude that very massive central galaxies are either extreme outliers with respect to their counterparts at \zf\ or, alternatively, have assembled most of their final stellar mass at $z<$\zf . Here, we probe the relative roles of star formation and mergers in growing massive galaxies in a late assembly scenario. For the former, we utilise empirical estimates of the star formation rates of massive galaxies as functions of redshift and stellar mass to evaluate if it is sufficient to grow the galaxy up to the stellar mass we observe today. For the latter, we utilise a more sophisticated semi-empirical model to evaluate the effectiveness of mergers in evolving the galaxy's stellar mass and size.

\subsection{Can massive galaxies grow solely through in-situ star formation?}

\begin{figure}[t!]
\centering
\includegraphics[width=0.5\textwidth]{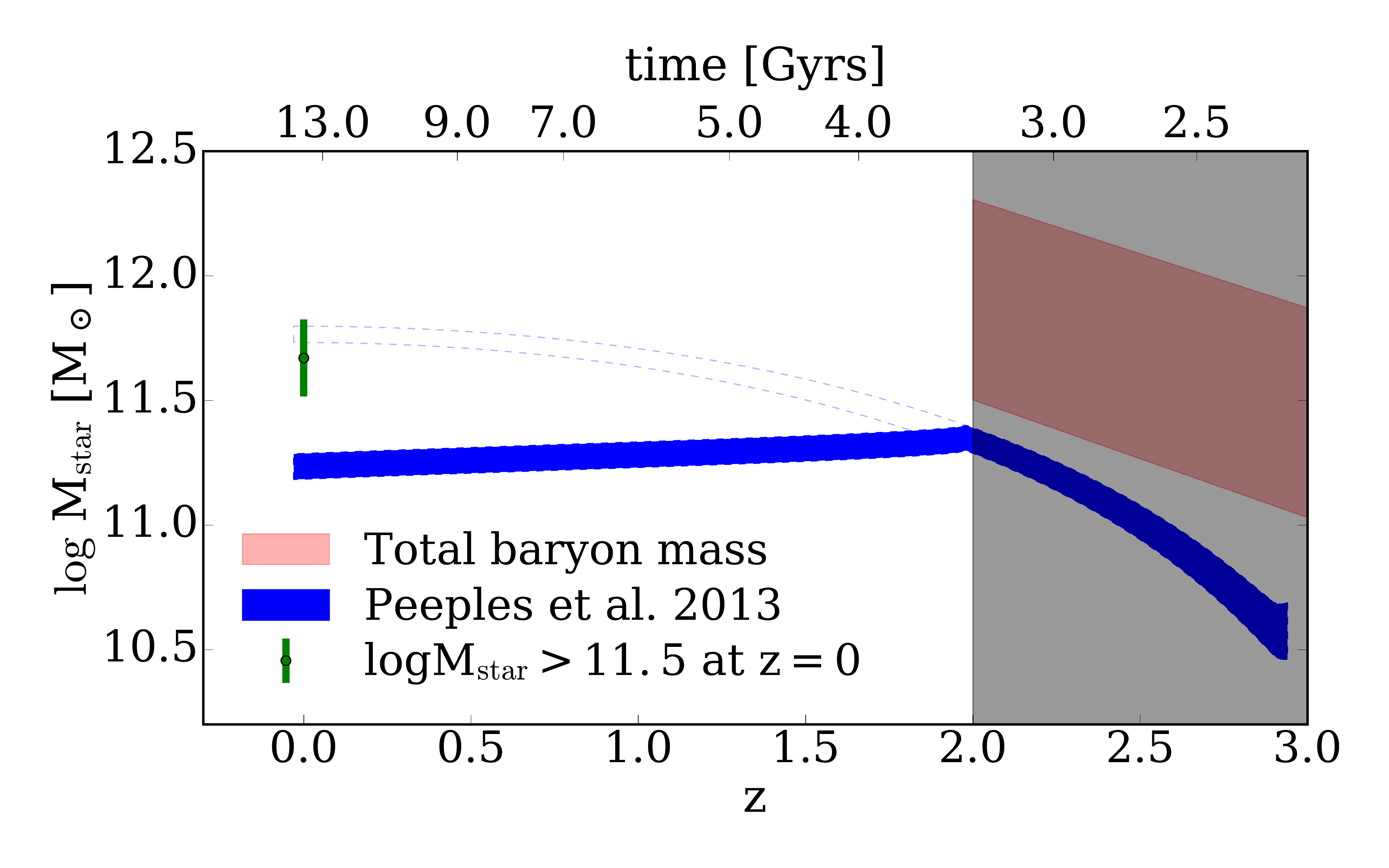}
\includegraphics[width=0.5\textwidth]{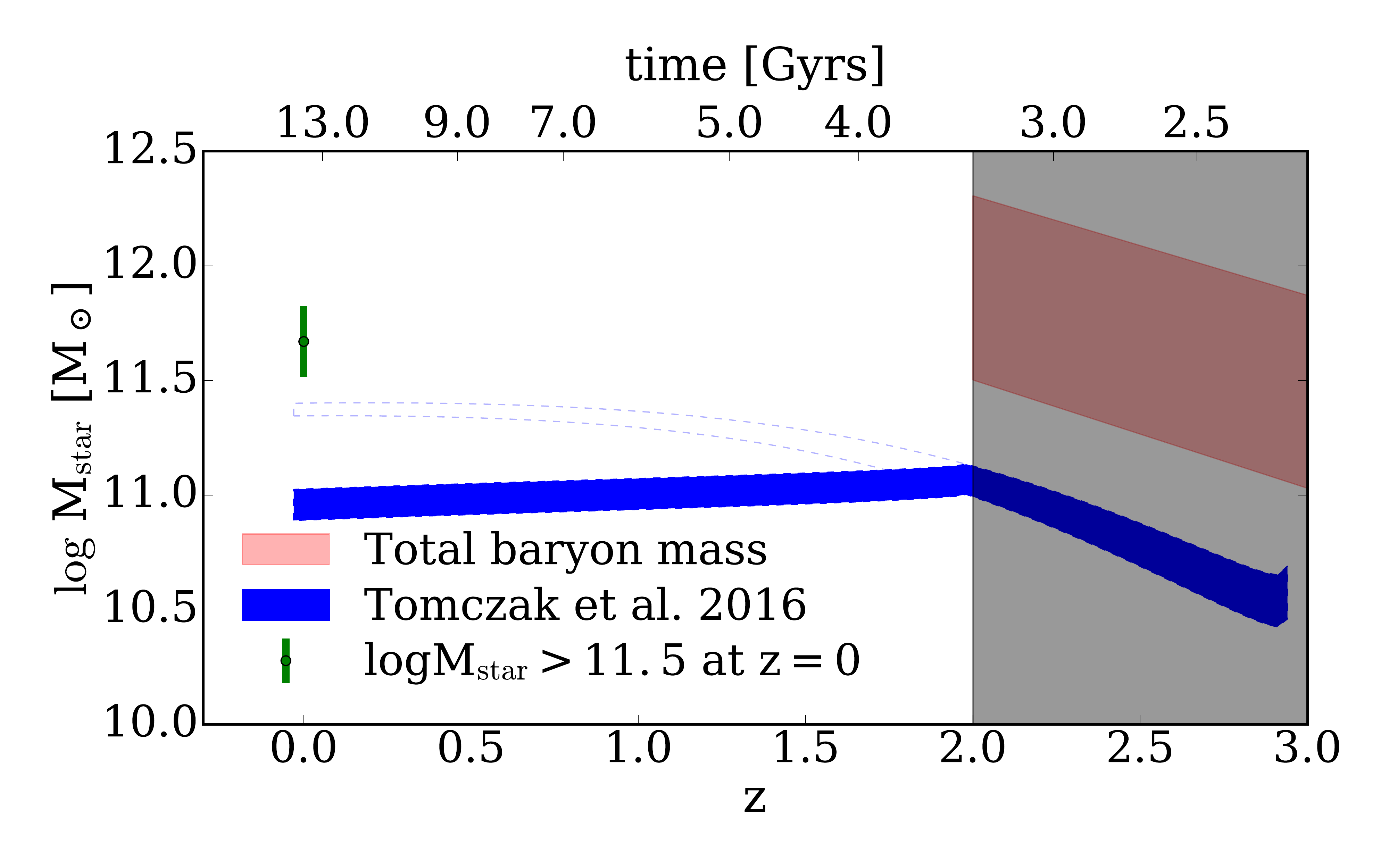}
\includegraphics[width=0.5\textwidth]{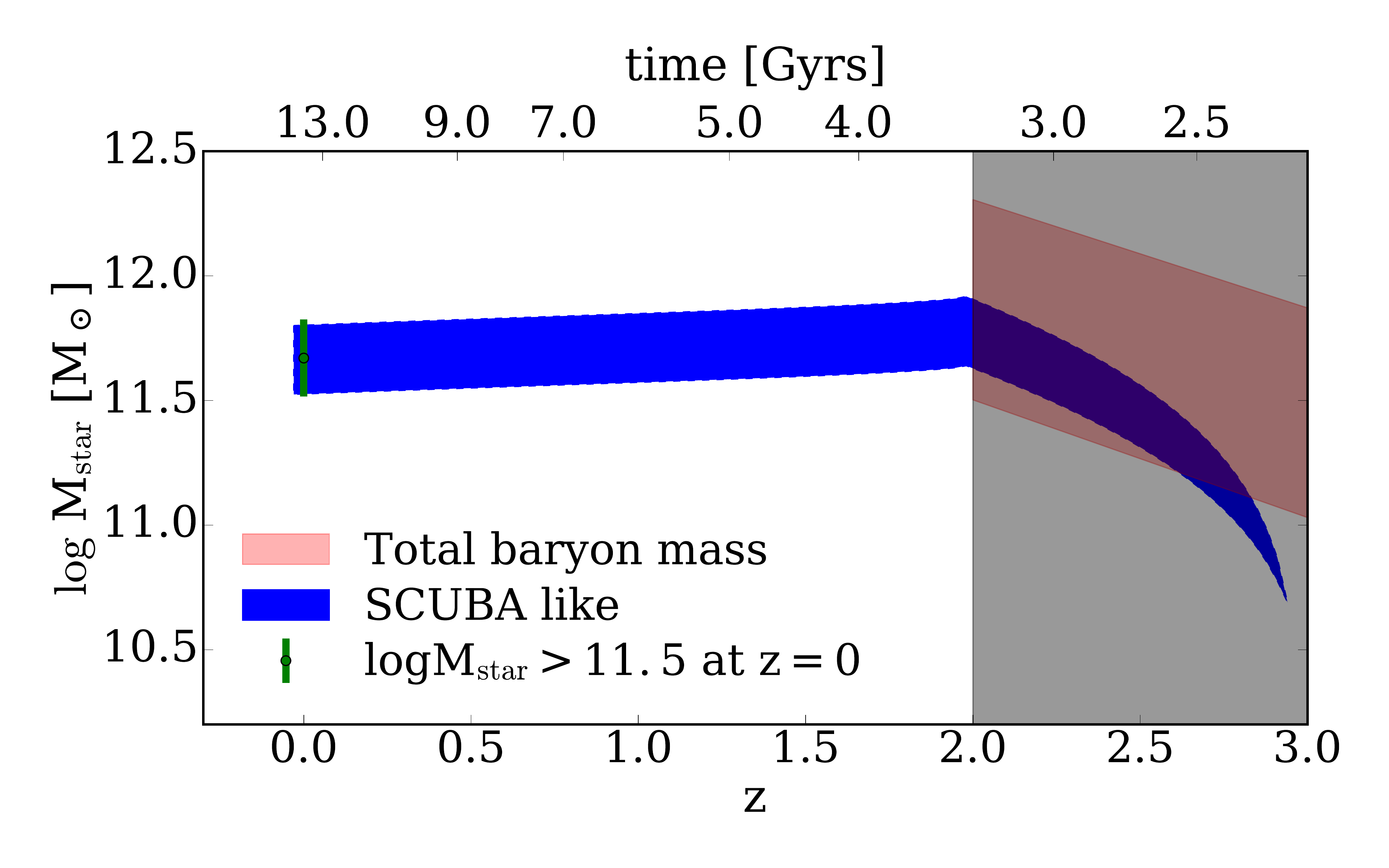}
\caption{The mean star formation track of galaxies selected with $\log\ ($\ms $)<11.5\ M_\odot$ at $z=0$ using the mass we use the empirical star formation rates of \protect\citet[][top panel]{Peeples2013} and \protect\citet[][middle panel]{Tomczak2016}. The band shows the systematic uncertainty in the \ms\ -\mh relation. The unfilled band shows the full evolutionary track from $z=3$ to $z=3$ and solid band shows the effects if the star formation is quenched at $z=2$. The bottom panel instead shows a band bracketing the mean evolutionary path assuming a constant star formation of $500$ and $1000\ M_\odot /yr$. For reference, we also show the total baryonic mass in the progenitor haloes as a red band between $z=2$ and $3$.}
\end{figure}

In this section, we utilise empirically-informed models to assess whether $\log\ ($\ms$)>11.5\ M_\odot$ galaxies at $z=0$ could have grown to their final stellar mass mostly through in-situ star formation, without the need for numerous mergers. We start by assuming that the progenitor galaxies are lying on the \ms\ -\mh\ and the specific star formation rate (sSFR)-\ms\ relations of typical main sequence galaxies at $z=$\zf. We take each of the $z=3$ progenitor haloes and assign to them a stellar mass using the \ms\ - \mh\ relation as described in section 2.1. For each of the galaxies, we evolve the stellar mass via redshift and mass dependent star formation rates but also accounting for the stellar mass loss of the evolving stellar population assuming a \cite{Chabrier2003} IMF. Specifically, we use the empirical star formation rates by \citet[][upper panel]{Peeples2013} and \citet[][lower panel]{Tomczak2016}, and use the \cite{Behroozi2013} prescription (their Equation 14) for the stellar mass loss. We assume that the galaxy can efficiently form stars, at least up to the maximum baryonic  content, assuming that the cold gas reservoir can be replenished via, e.g., cold flows \citep{Dekel2009}.

Figure 6 shows the mean evolutionary track of our sub-sample of galaxies with $\log\ ($\ms$)>11.5\ M_\odot$ at $z=0$. The blue band represents the systematic uncertainty in the \ms\ -\mh\ relation (model I and II as described in section 2.1). We find that assuming these galaxies remain on the star forming main sequence until $z=0$, the \cite{Peeples2013} star formation rate (top panel) can fully account for the observed mass measured in the local universe (dotted lines). The \cite{Tomczak2016} star formation (bottom panel) predicts a final stellar mass which is a factor of two to three lower. However, systems grown via a very prolonged star formation episode would be inconsistent with the observed ages (and colours) of very massive, central galaxies at $z=0$. As discussed in section 2.3, \cite{McDermid2015} claim that from their spectral fitting, 90\% of stellar mass in galaxies with $\log\ ($\ms$)>11.3-11.5\ M_\odot$ was formed at $z>2$ with a slight dependence on environment. Also, as mentioned in section 1, very massive galaxies have enhanced alpha element abundances relative to iron. In a closed box model, values of $[\alpha /Fe]\sim 0.17$, which are consistent with those measured in BCGs \citep[eg,][]{Oliva-Altamirano2015}, imply time scales of $<0.5-1$ Gyrs, according to the approximation
\begin{equation}
[\alpha/Fe]\sim \frac{1}{5}-\frac{1}{6}\times \Delta t (\text{Gyrs})
\end{equation}
from \citet{Thomas2005} and \cite{Citro2016}, which is in broad agreement with other studies of chemical evolution of massive galaxies \citep[e.g.,][]{Granato2004,Conroy2014}. If star formation is longer than $\sim 1$ Gyrs then the alpha element enhancement will be rapidly washed out by type Ia supernovae. The filled blue bands in Figure 6 show the predicted stellar mass evolution in the hypothesis that massive galaxies form by $z\simeq 3$ and quench by $z\simeq 2$ satisfying the conditions that the burst of star formation is limited to $\Delta t\leqslant 1$ Gyrs and that most of the stars are formed by $z\simeq 2$. It can be seen that, in the assumption that no new star formation takes place at $z<2$, the resultant stellar mass is at most $\log\ ($\ms$)\lesssim 11.2$ which is a factor of at least three less that what is measured at $z=0$. 

Alternatively, massive galaxies could be, as mentioned earlier, extreme outliers in both the \ms\ -\mh\ and/or sSFR-\ms\ relations. For example, if our progenitors were closer to SCUBA/ULIRGs with star formation rates up to 500-1000 $M_\odot/yr$, they could easily reach the stellar mass measured at z=0 in less than $0.5-1$ Gyrs. This is illustrated in the bottom panel of Figure 6 which shows with a blue band the mean mass growth of galaxies assuming a constant star formation of $500$ and $1000 M_\odot/yr$. To maintain such a high star formation rates, progenitor galaxies would need to turn about all of their initial baryonic content in the host dark matter halo (red regions) into stars. Additionally, this scenario would not allow for the observed, significant mass loss by stellar and/or AGN winds seen in a number of ULIRGs \citep[e.g.,][]{Smail2003,Swinbank2005}.

\subsection{Can mergers drive mass evolution of massive galaxies?}

We now turn to explore the possibility that (mainly dry, minor) mergers are the main driver behind the mass and size evolution of massive, central galaxies. To this purpose, we utilise a more sophisticated, state of the art semi-empirical model (SEM). A full description of our SEM is given by \cite{Shankar2014a,Shankar2015}, but we also provide a brief overview here. 

The SEM is constructed on top of dark matter merger trees extracted from the Millennium simulation \citep{Springel2005}. At the formation redshift, the main progenitors haloes are `populated' with star forming, disc galaxies which, by construction, follow empirical relations for stellar mass \citep{Moster2013}, gas fraction \citep{Stewart2009}, disc radius \citep{Shen2003}, and star formation rate \citep{Peeples2013}. At each timestep, the main progenitor galaxies are re-initialised using these empirical relations thus bypassing the need to model the full, complex and still unclear aspects of galaxy formation, such as cooling and feedback.

When sub-haloes infall into the halo hosting the massive galaxy, we assign to them a satellite galaxy with stellar mass given by abundance matching relations at the redshift of infall. The structural properties of this new satellite are equal to a previously simulated, random central galaxy extracted from the model with equal stellar mass at the the redshift of infall. The new satellite galaxy is allowed to orbit for a dynamical friction timescale. Over this time, the satellite can grow in stellar mass and size according to its available gas and star formation rate at infall. If a merger between the central and satellite galaxy occurs, the stellar mass and gas mass of the satellite are added to the bulge and disc of the central galaxy, respectively. The new radius of the bulge is calculated by conserving the sum of the binding energies and the mutual orbital energy of the two merging galaxies \citep{Cole2000}.

\begin{figure}[h!]
\centering
\includegraphics[width=0.45\textwidth]{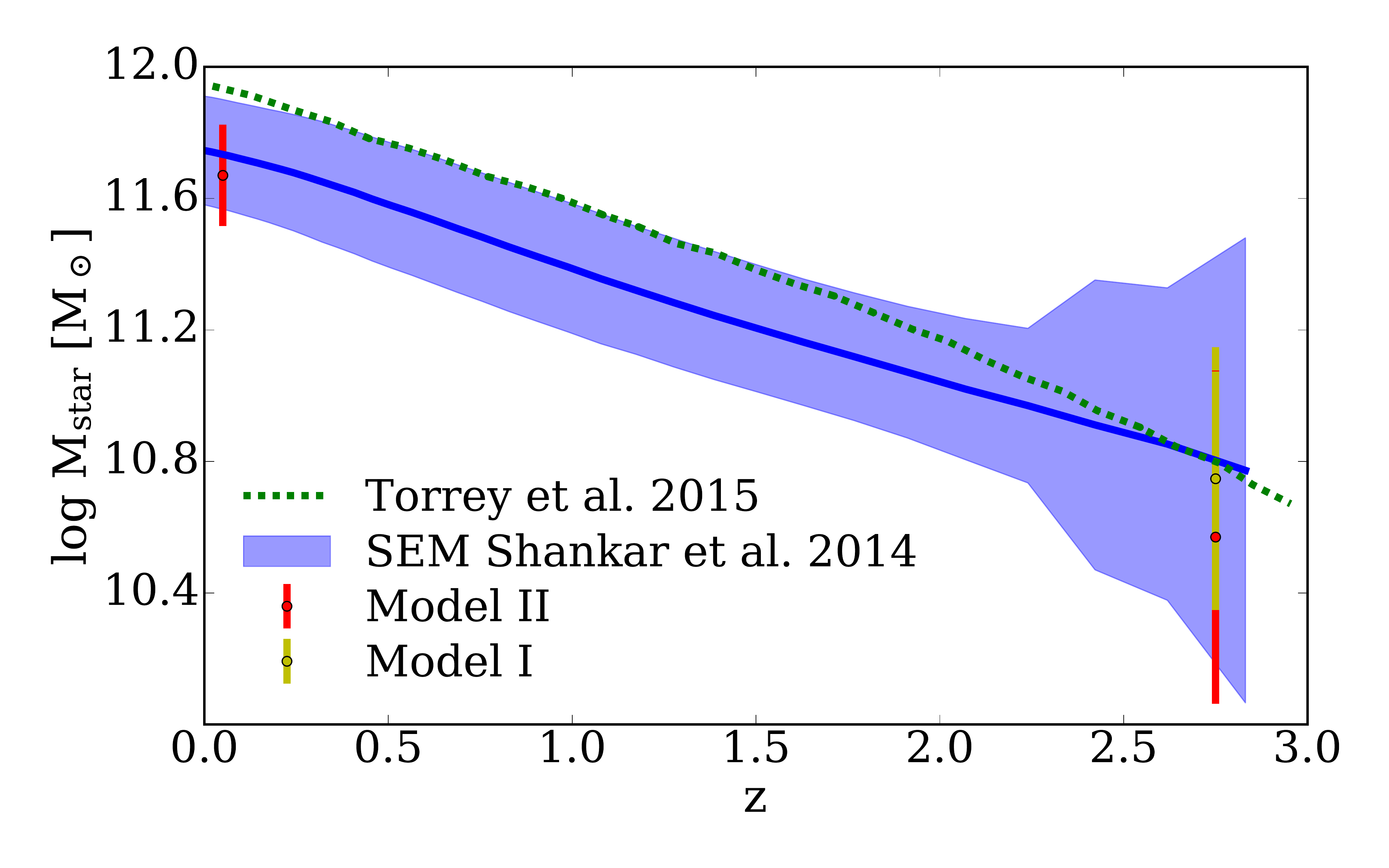}
\caption{The mean evolution of the stellar mass of a galaxies with $log\ ($\ms$)>11.5$ at $z=0$ evolved using the SEM in \protect\cite{Shankar2014a}. The shaded region shows the statistic dispersion in the galaxies' evolutionary histories. The error bars show the predictions from abundance matching for galaxies with $\log\ ($\ms$)>11.5\ M_\odot$ where model I and II are shown in yellow and red respectively. For reference, we also show the the mean evolution in the mass of the most massive galaxies in the illustris simulation \citep{Torrey2015}.}
\end{figure}

Figure 7 shows the mean mass evolution of the most massive galaxies evolved using our SEM. The shaded region represents the $1\sigma$ dispersion in the mean stellar mass at any redshift. We find that the merger-driven SEM is capable of reproducing the median mass evolution of the most massive galaxies. This result confirms, and extends, what was found by \cite{Shankar2015} from the evolution of brightest cluster galaxies from $z\sim 1$ to $0$. This result is also in agreement with previous works mainly based on high-resolution N-body simulations which showed that the history of the centre of  clusters is highly affected by frequent mergers. For example, \cite{Gao2004} discussed that a typical massive brightest cluster galaxy should have undergone a significant number of merging events even at $z<1$. More recently this has been further discussed and confirmed by \cite{Laporte2015}. As shown in Figure 7, the results from the semi-empirical model are also in agreement with the full, cosmological hydro-simulations of \cite{Torrey2015}.

\subsection{Can mergers drive size evolution of massive galaxies?}

Additional hints come from the size evolution of the central galaxies. Figure 8 shows a comparison between the size evolution of massive ETGs evolved using our SEM to the observations of \cite{Bernardi2014} in the local universe. We also plot the observed size of the putative progenitor at $z=2.75$ by using the stellar masses predicted by the SEM and using the size-mass relation for disc galaxies found by \cite{VanderWel2014}. The shaded region in Figure 8 shows the systematic uncertainty in the predicted mean size evolution caused by allowing for some stellar stripping at the level suggested by observations \citep[see][for full details]{Cattaneo2011,Shankar2014a}. We find that, irrespective of the exact level of (stellar) stripping, the SEM is fully capable of reproducing the observed mean size evolution of the most massive galaxies, in line with the conclusions of \cite{Shankar2015} at $z<1$.

\begin{figure}

\centering
\includegraphics[width=0.5\textwidth]{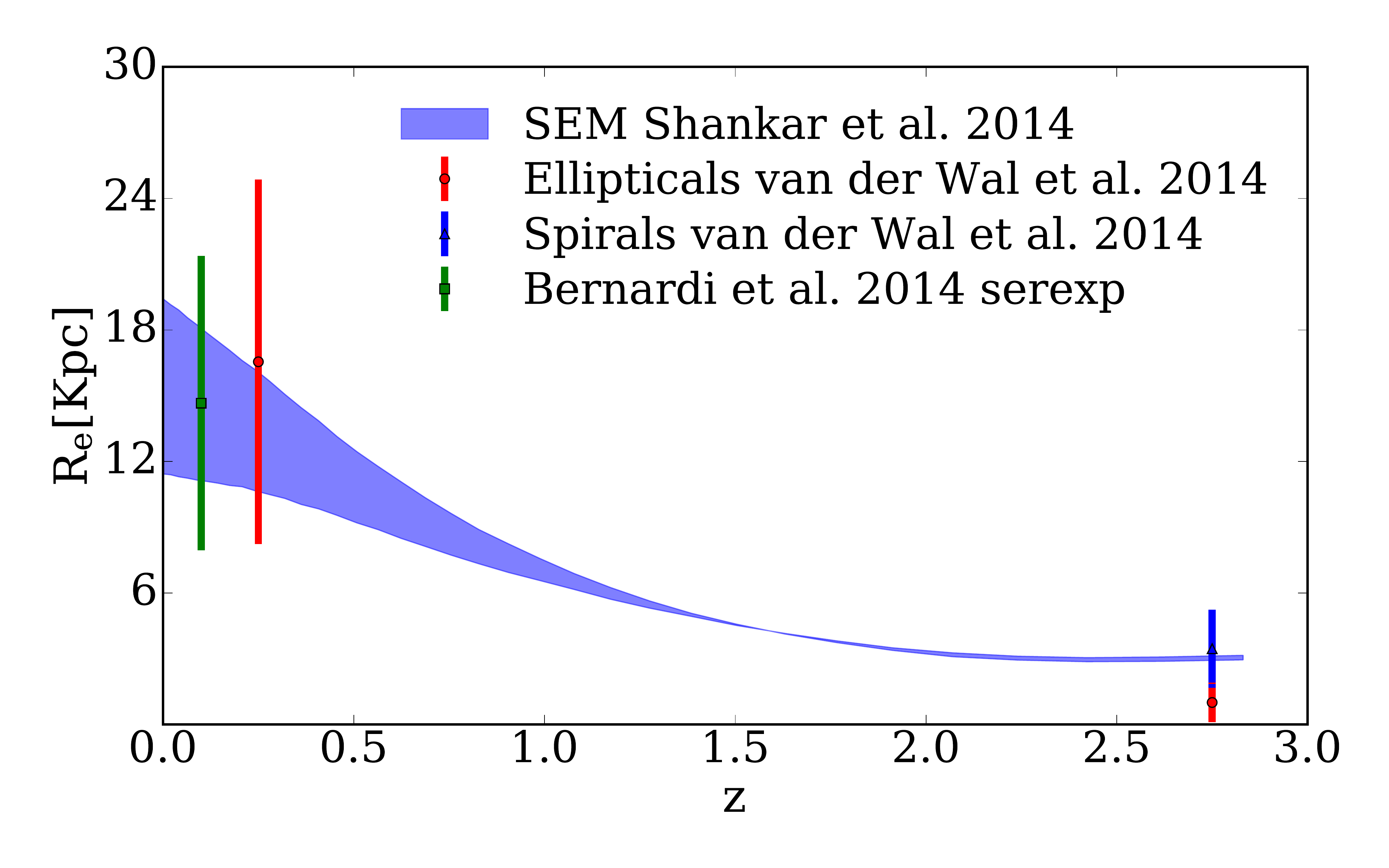}
\caption{The median evolution of the effective radius of a galaxies with $\log\ ($\ms$)>11.5$ at $z=0$ evolved using the SEM in \protect\cite{Shankar2014a}. The shaded area represents the systematic uncertainty of the model when gas dispersion included.}
\end{figure}

\subsection{Size growth from quasar mode feedback}

The extreme assumption of a very efficient collapse in which the majority of initial baryons are converted into stars, as discussed in section 4.1, would clearly not allow for any size growth from \zf$>2$\ to $z=0$. Even in the puffing up scenario proposed by \cite{Fan2008} and discussed in section 1, a significant fraction of the baryons must be lost via stellar winds and/or quasar mode feedback at \zf\ for the galaxy to react quasi-adiabatically and expand. Analytic arguments by \cite{Fan2010}, backed up by numerical simulations by \cite{Ragone-Figueroa2011}, have shown that the increase in effective radius is roughly proportional to the amount of mass lost. This would imply that to allow for a factor of at least three increase in mean size since $z\sim 3$, as observationally inferred by \cite{VanderWel2014}, $\gtrsim 70\%$ of the total initial baryons must be expelled from the galaxy \citep{Fan2010}. A mass loss of $\gtrsim 70\%$ at $z>2$ would place the residual baryonic mass in the progenitor halo significantly below the descendants' stellar mass at $z=0$. Thus, even in an efficient quasar-feedback scenario, our progenitor-descendant evolutionary tracks would still require substantial late assembly of stellar mass via, e.g., mergers.

\subsection{No size growth for very massive ETGs?}

Interestingly, there is mounting observational evidence for very massive galaxies at the centre of high redshift clusters that sit already on the local size-mass relation of early-type galaxies \citep{Strazzullo2013,Delaye2014, Newman2014}. This might be in support of the hypothesis of very rapid and efficient bursts of star formation. However, even if massive galaxies are indeed true outliers with respect to the general population of central galaxies with \mh $\sim10^{12} M_\odot$ at \zf$\geq 2$, they would still require an initial star formation burst capable of converting $\approx 100\%$ of the initial baryons into stars. Thus, one clear observational prediction we can make is that the progenitors of massive galaxies should either be moderately massive and  compact, or very massive and extended.

\section{Conclusions}

In this work, we have set tighter constrains on the assembly and evolution of massive, central galaxies. We utilise a catalogue of dark matter haloes created from the Bolshoi simulation. We populate these haloes with a stellar mass using recent rendition of the stellar mass to halo mass relation by \cite{Kravtsov2014} and \cite{Shankar2014b} at $z=0$ and select haloes with $\log\ ($\ms$)> 11.5\ M_\odot$. We then trace host haloes back to the putative formation epoch, \zf$=2-4$, as inferred from the stellar ages of massive ETGs. At this epoch, we estimate the total mass in baryons within the halo from the baryon fraction. We find that the stellar mass of the ETG in the local universe is comparable to, if not higher than, the total baryonic mass contained within the progenitor halo. From this comparison, we draw the following important conclusions.

\begin{enumerate}

\item In-situ formation: For these massive galaxies to have fully assembled at the formation epoch, the efficiency of converting baryons into stars needs to be extremely high if not 100\%. We also show that this assembly scenario would lead to all ETGs being extreme outliers with respect to what is predicted by abundance matching at \zf.

\item Size: Even when assuming an extremely efficient star formation at \zf, the galaxy would not be allowed any size growth since the formation epoch. Even an in-situ expansion would in fact require a mass loss of $\geq 70\%$ of the initial baryon content to be sufficiently efficient. Thus, in a strictly monolithic scenario, progenitors of massive galaxies should already be extended systems at their formation epoch. Measurements of the structure of massive galaxies in massive haloes will be critical to assess this possibility.

\item Late assembly: Star formation could contribute to the stellar mass growth of the progenitors of massive galaxies, but cannot explain their full evolution. We show through state-of-the-art, cosmological, semi-empirical models that massive galaxies could have indeed assembled most of their final mass via late mergers and be consistent with available data on their size evolution. It remains to be seen the impact of mergers on other (tight) galaxy scaling relations involving velocity dispersion \citep[e.g.,][and references therein]{Bernardi2011b, Bernardi2011a,Shankar2016}.

\end{enumerate}

More secure and statistically relevant measurements of the stellar mass and structure of high redshift brightest cluster galaxies will be of key relevance to discern between merger scenarios and extremely efficient starbursts events. 

\section*{Acknowledgements}
The authors warmly thank an anonymous referee for comments that helped improve the presentation of our results and Vincent Bouillot, Ravi Sheth, Andrea Lapi, Luigi Danese and Ignacio Trujillo for useful discussions.

\bibliographystyle{mnras} 
\bibliography{library.bib}

\begin{thebibliography}{}
\makeatletter
\relax
\def\mn@urlcharsother{\let\do\@makeother \do\$\do\&\do\#\do\^\do\_\do\%\do\~}
\def\mn@doi{\begingroup\mn@urlcharsother \@ifnextchar [ {\mn@doi@}
  {\mn@doi@[]}}
\def\mn@doi@[#1]#2{\def\@tempa{#1}\ifx\@tempa\@empty \href
  {http://dx.doi.org/#2} {doi:#2}\else \href {http://dx.doi.org/#2} {#1}\fi
  \endgroup}
\def\mn@eprint#1#2{\mn@eprint@#1:#2::\@nil}
\def\mn@eprint@arXiv#1{\href {http://arxiv.org/abs/#1} {{\tt arXiv:#1}}}
\def\mn@eprint@dblp#1{\href {http://dblp.uni-trier.de/rec/bibtex/#1.xml}
  {dblp:#1}}
\def\mn@eprint@#1:#2:#3:#4\@nil{\def\@tempa {#1}\def\@tempb {#2}\def\@tempc
  {#3}\ifx \@tempc \@empty \let \@tempc \@tempb \let \@tempb \@tempa \fi \ifx
  \@tempb \@empty \def\@tempb {arXiv}\fi \@ifundefined
  {mn@eprint@\@tempb}{\@tempb:\@tempc}{\expandafter \expandafter \csname
  mn@eprint@\@tempb\endcsname \expandafter{\@tempc}}}

\bibitem[\protect\citeauthoryear{{Ahn} et~al.,}{{Ahn} et~al.}{2014}]{Ahn2014}
{Ahn} C.~P.,  et~al., 2014, \mn@doi [\apjs] {10.1088/0067-0049/211/2/17}, \href
  {http://adsabs.harvard.edu/abs/2014ApJS..211...17A} {211, 17}

\bibitem[\protect\citeauthoryear{{Aversa}, {Lapi}, {de Zotti}, {Shankar}  \&
  {Danese}}{{Aversa} et~al.}{2015}]{Aversa2015}
{Aversa} R.,  {Lapi} A.,  {de Zotti} G.,  {Shankar} F.,   {Danese} L.,  2015,
  \mn@doi [\apj] {10.1088/0004-637X/810/1/74}, \href
  {http://adsabs.harvard.edu/abs/2015ApJ...810...74A} {810, 74}

\bibitem[\protect\citeauthoryear{Baldry, Glazebrook, Brinkmann, Ivezi\'{c},
  Lupton, Nichol  \& Szalay}{Baldry et~al.}{2004}]{Baldry2004}
Baldry I.~K.,  Glazebrook K.,  Brinkmann J.,  Ivezi\'{c} v.,  Lupton R.~H.,
  Nichol R.~C.,   Szalay A.~S.,  2004, \mn@doi [\apj] {10.1086/380092}, 600,
  681

\bibitem[\protect\citeauthoryear{{Baugh}}{{Baugh}}{2006}]{Baugh2006}
{Baugh} C.~M.,  2006, \mn@doi [Reports on Progress in Physics]
  {10.1088/0034-4885/69/12/R02}, \href
  {http://adsabs.harvard.edu/abs/2006RPPh...69.3101B} {69, 3101}

\bibitem[\protect\citeauthoryear{{Behroozi}, {Wechsler}  \&
  {Conroy}}{{Behroozi} et~al.}{2013}]{Behroozi2013}
{Behroozi} P.~S.,  {Wechsler} R.~H.,   {Conroy} C.,  2013, \mn@doi [\apj]
  {10.1088/0004-637X/770/1/57}, \href
  {http://adsabs.harvard.edu/abs/2013ApJ...770...57B} {770, 57}

\bibitem[\protect\citeauthoryear{{Benson}}{{Benson}}{2012}]{Benson2012}
{Benson} A.~J.,  2012, \mn@doi [\na] {10.1016/j.newast.2011.07.004}, \href
  {http://adsabs.harvard.edu/abs/2012NewA...17..175B} {17, 175}

\bibitem[\protect\citeauthoryear{{Bernardi}, {Roche}, {Shankar}  \&
  {Sheth}}{{Bernardi} et~al.}{2011a}]{Bernardi2011b}
{Bernardi} M.,  {Roche} N.,  {Shankar} F.,   {Sheth} R.~K.,  2011a, \mn@doi
  [\mnras] {10.1111/j.1745-3933.2010.00982.x}, \href
  {http://adsabs.harvard.edu/abs/2011MNRAS.412L...6B} {412, L6}

\bibitem[\protect\citeauthoryear{{Bernardi}, {Roche}, {Shankar}  \&
  {Sheth}}{{Bernardi} et~al.}{2011b}]{Bernardi2011a}
{Bernardi} M.,  {Roche} N.,  {Shankar} F.,   {Sheth} R.~K.,  2011b, \mn@doi
  [\mnras] {10.1111/j.1365-2966.2010.17984.x}, \href
  {http://adsabs.harvard.edu/abs/2011MNRAS.412..684B} {412, 684}

\bibitem[\protect\citeauthoryear{{Bernardi}, {Meert}, {Sheth}, {Vikram},
  {Huertas-Company}, {Mei}  \& {Shankar}}{{Bernardi}
  et~al.}{2013}]{Bernardi2013}
{Bernardi} M.,  {Meert} A.,  {Sheth} R.~K.,  {Vikram} V.,  {Huertas-Company}
  M.,  {Mei} S.,   {Shankar} F.,  2013, \mn@doi [\mnras]
  {10.1093/mnras/stt1607}, \href
  {http://adsabs.harvard.edu/abs/2013MNRAS.436..697B} {436, 697}

\bibitem[\protect\citeauthoryear{{Bernardi}, {Meert}, {Vikram},
  {Huertas-Company}, {Mei}, {Shankar}  \& {Sheth}}{{Bernardi}
  et~al.}{2014}]{Bernardi2014}
{Bernardi} M.,  {Meert} A.,  {Vikram} V.,  {Huertas-Company} M.,  {Mei} S.,
  {Shankar} F.,   {Sheth} R.~K.,  2014, \mn@doi [\mnras]
  {10.1093/mnras/stu1106}, \href
  {http://adsabs.harvard.edu/abs/2014MNRAS.443..874B} {443, 874}

\bibitem[\protect\citeauthoryear{{Bernardi}, {Meert}, {Sheth}, {Fischer},
  {Huertas-Company}, {Maraston}, {Shankar}  \& {Vikram}}{{Bernardi}
  et~al.}{2016a}]{Bernardi2016b}
{Bernardi} M.,  {Meert} A.,  {Sheth} R.~K.,  {Fischer} J.-L.,
  {Huertas-Company} M.,  {Maraston} C.,  {Shankar} F.,   {Vikram} V.,  2016a,
  preprint, \href {http://adsabs.harvard.edu/abs/2016arXiv160401036B} {}
  (\mn@eprint {arXiv} {1604.01036})

\bibitem[\protect\citeauthoryear{{Bernardi}, {Meert}, {Sheth},
  {Huertas-Company}, {Maraston}, {Shankar}  \& {Vikram}}{{Bernardi}
  et~al.}{2016b}]{Bernardi2016}
{Bernardi} M.,  {Meert} A.,  {Sheth} R.~K.,  {Huertas-Company} M.,  {Maraston}
  C.,  {Shankar} F.,   {Vikram} V.,  2016b, \mn@doi [\mnras]
  {10.1093/mnras/stv2487}, \href
  {http://adsabs.harvard.edu/abs/2016MNRAS.455.4122B} {455, 4122}

\bibitem[\protect\citeauthoryear{{Bond}, {Cole}, {Efstathiou}  \&
  {Kaiser}}{{Bond} et~al.}{1991}]{Bond1991}
{Bond} J.~R.,  {Cole} S.,  {Efstathiou} G.,   {Kaiser} N.,  1991, \mn@doi
  [\apj] {10.1086/170520}, \href
  {http://adsabs.harvard.edu/abs/1991ApJ...379..440B} {379, 440}

\bibitem[\protect\citeauthoryear{{Bruzual} \& {Charlot}}{{Bruzual} \&
  {Charlot}}{2003}]{Bruzual2003}
{Bruzual} G.,  {Charlot} S.,  2003, \mn@doi [\mnras]
  {10.1046/j.1365-8711.2003.06897.x}, \href
  {http://adsabs.harvard.edu/abs/2003MNRAS.344.1000B} {344, 1000}

\bibitem[\protect\citeauthoryear{{Cappellari} et~al.,}{{Cappellari}
  et~al.}{2013}]{Cappellari2013}
{Cappellari} M.,  et~al., 2013, \mn@doi [\mnras] {10.1093/mnras/stt644}, \href
  {http://adsabs.harvard.edu/abs/2013MNRAS.432.1862C} {432, 1862}

\bibitem[\protect\citeauthoryear{{Carollo} et~al.,}{{Carollo}
  et~al.}{2013}]{Carollo2013}
{Carollo} C.~M.,  et~al., 2013, \mn@doi [\apj] {10.1088/0004-637X/773/2/112},
  \href {http://adsabs.harvard.edu/abs/2013ApJ...773..112C} {773, 112}

\bibitem[\protect\citeauthoryear{{Cassata} et~al.,}{{Cassata}
  et~al.}{2008}]{Cassata2008}
{Cassata} P.,  et~al., 2008, \mn@doi [\aap] {10.1051/0004-6361:200809881},
  \href {http://adsabs.harvard.edu/abs/2008A%26A...483L..39C} {483, L39}

\bibitem[\protect\citeauthoryear{{Cattaneo}, {Mamon}, {Warnick}  \&
  {Knebe}}{{Cattaneo} et~al.}{2011}]{Cattaneo2011}
{Cattaneo} A.,  {Mamon} G.~A.,  {Warnick} K.,   {Knebe} A.,  2011, \mn@doi
  [\aap] {10.1051/0004-6361/201015780}, \href
  {http://adsabs.harvard.edu/abs/2011A%26A...533A...5C} {533, A5}

\bibitem[\protect\citeauthoryear{{Chabrier}}{{Chabrier}}{2003}]{Chabrier2003}
{Chabrier} G.,  2003, \mn@doi [\pasp] {10.1086/376392}, \href
  {http://adsabs.harvard.edu/abs/2003PASP..115..763C} {115, 763}

\bibitem[\protect\citeauthoryear{{Chapman}, {Blain}, {Smail}  \&
  {Ivison}}{{Chapman} et~al.}{2005}]{Chapman2005}
{Chapman} S.~C.,  {Blain} A.~W.,  {Smail} I.,   {Ivison} R.~J.,  2005, \mn@doi
  [\apj] {10.1086/428082}, \href
  {http://adsabs.harvard.edu/abs/2005ApJ...622..772C} {622, 772}

\bibitem[\protect\citeauthoryear{{Chiosi}, {Merlin}  \& {Piovan}}{{Chiosi}
  et~al.}{2012}]{Chiosi2012}
{Chiosi} C.,  {Merlin} E.,   {Piovan} L.,  2012, preprint, \href
  {http://adsabs.harvard.edu/abs/2012arXiv1206.2532C} {} (\mn@eprint {arXiv}
  {1206.2532})

\bibitem[\protect\citeauthoryear{{Citro}, {Pozzetti}, {Moresco}  \&
  {Cimatti}}{{Citro} et~al.}{2016}]{Citro2016}
{Citro} A.,  {Pozzetti} L.,  {Moresco} M.,   {Cimatti} A.,  2016, preprint,
  \href {http://adsabs.harvard.edu/abs/2016arXiv160407826C} {} (\mn@eprint
  {arXiv} {1604.07826})

\bibitem[\protect\citeauthoryear{{Clauwens}, {Franx}  \& {Schaye}}{{Clauwens}
  et~al.}{2016}]{Clauwens2016}
{Clauwens} B.,  {Franx} M.,   {Schaye} J.,  2016, preprint, \href
  {http://adsabs.harvard.edu/abs/2016arXiv160500009C} {} (\mn@eprint {arXiv}
  {1605.00009})

\bibitem[\protect\citeauthoryear{{Cole}, {Lacey}, {Baugh}  \& {Frenk}}{{Cole}
  et~al.}{2000}]{Cole2000}
{Cole} S.,  {Lacey} C.~G.,  {Baugh} C.~M.,   {Frenk} C.~S.,  2000, \mn@doi
  [\mnras] {10.1046/j.1365-8711.2000.03879.x}, \href
  {http://adsabs.harvard.edu/abs/2000MNRAS.319..168C} {319, 168}

\bibitem[\protect\citeauthoryear{{Conroy}, {Graves}  \& {van Dokkum}}{{Conroy}
  et~al.}{2014}]{Conroy2014}
{Conroy} C.,  {Graves} G.~J.,   {van Dokkum} P.~G.,  2014, \mn@doi [\apj]
  {10.1088/0004-637X/780/1/33}, \href
  {http://adsabs.harvard.edu/abs/2014ApJ...780...33C} {780, 33}

\bibitem[\protect\citeauthoryear{{Conselice}}{{Conselice}}{2014}]{Conselice2014}
{Conselice} C.~J.,  2014, \mn@doi [\araa]
  {10.1146/annurev-astro-081913-040037}, \href
  {http://adsabs.harvard.edu/abs/2014ARA%26A..52..291C} {52, 291}

\bibitem[\protect\citeauthoryear{{Crain}, {Eke}, {Frenk}, {Jenkins},
  {McCarthy}, {Navarro}  \& {Pearce}}{{Crain} et~al.}{2007}]{Crain2007}
{Crain} R.~A.,  {Eke} V.~R.,  {Frenk} C.~S.,  {Jenkins} A.,  {McCarthy} I.~G.,
  {Navarro} J.~F.,   {Pearce} F.~R.,  2007, \mn@doi [\mnras]
  {10.1111/j.1365-2966.2007.11598.x}, \href
  {http://adsabs.harvard.edu/abs/2007MNRAS.377...41C} {377, 41}

\bibitem[\protect\citeauthoryear{{D'Souza}, {Vegetti}  \&
  {Kauffmann}}{{D'Souza} et~al.}{2015}]{D'Souz2015}
{D'Souza} R.,  {Vegetti} S.,   {Kauffmann} G.,  2015, \mn@doi [\mnras]
  {10.1093/mnras/stv2234}, \href
  {http://adsabs.harvard.edu/abs/2015MNRAS.454.4027D} {454, 4027}

\bibitem[\protect\citeauthoryear{{Damjanov} et~al.,}{{Damjanov}
  et~al.}{2009}]{Damjanov2009}
{Damjanov} I.,  et~al., 2009, \mn@doi [\apj] {10.1088/0004-637X/695/1/101},
  \href {http://adsabs.harvard.edu/abs/2009ApJ...695..101D} {695, 101}

\bibitem[\protect\citeauthoryear{{Dawson} et~al.,}{{Dawson}
  et~al.}{2013}]{BOSS2013}
{Dawson} K.~S.,  et~al., 2013, \mn@doi [\aj] {10.1088/0004-6256/145/1/10},
  \href {http://adsabs.harvard.edu/abs/2013AJ....145...10D} {145, 10}

\bibitem[\protect\citeauthoryear{{De Lucia}, {Fontanot}, {Wilman}  \&
  {Monaco}}{{De Lucia} et~al.}{2011}]{DeLucia2011}
{De Lucia} G.,  {Fontanot} F.,  {Wilman} D.,   {Monaco} P.,  2011, \mn@doi
  [\mnras] {10.1111/j.1365-2966.2011.18475.x}, \href
  {http://adsabs.harvard.edu/abs/2011MNRAS.414.1439D} {414, 1439}

\bibitem[\protect\citeauthoryear{{Dekel} et~al.,}{{Dekel}
  et~al.}{2009}]{Dekel2009}
{Dekel} A.,  et~al., 2009, \mn@doi [nat] {10.1038/nature07648}, \href
  {http://adsabs.harvard.edu/abs/2009Natur.457..451D} {457, 451}

\bibitem[\protect\citeauthoryear{{Delaye} et~al.,}{{Delaye}
  et~al.}{2014}]{Delaye2014}
{Delaye} L.,  et~al., 2014, \mn@doi [\mnras] {10.1093/mnras/stu496}, \href
  {http://adsabs.harvard.edu/abs/2014MNRAS.441..203D} {441, 203}

\bibitem[\protect\citeauthoryear{{Duc}, {Cuillandre}  \& {Karabal}}{{Duc}
  et~al.}{2015}]{Duc2015}
{Duc} P.-A.,  {Cuillandre} J.-C.,   {Karabal} 2015, \mn@doi [\mnras]
  {10.1093/mnras/stu2019}, \href
  {http://adsabs.harvard.edu/abs/2015MNRAS.446..120D} {446, 120}

\bibitem[\protect\citeauthoryear{{Fan}, {Lapi}, {De Zotti}  \& {Danese}}{{Fan}
  et~al.}{2008}]{Fan2008}
{Fan} L.,  {Lapi} A.,  {De Zotti} G.,   {Danese} L.,  2008, \mn@doi [\apjl]
  {10.1086/595784}, \href {http://adsabs.harvard.edu/abs/2008ApJ...689L.101F}
  {689, L101}

\bibitem[\protect\citeauthoryear{{Fan}, {Lapi}, {Bressan}, {Bernardi}, {De
  Zotti}  \& {Danese}}{{Fan} et~al.}{2010}]{Fan2010}
{Fan} L.,  {Lapi} A.,  {Bressan} A.,  {Bernardi} M.,  {De Zotti} G.,   {Danese}
  L.,  2010, \mn@doi [\apj] {10.1088/0004-637X/718/2/1460}, \href
  {http://adsabs.harvard.edu/abs/2010ApJ...718.1460F} {718, 1460}

\bibitem[\protect\citeauthoryear{{Gao}, {Loeb}, {Peebles}, {White}  \&
  {Jenkins}}{{Gao} et~al.}{2004}]{Gao2004}
{Gao} L.,  {Loeb} A.,  {Peebles} P.~J.~E.,  {White} S.~D.~M.,   {Jenkins} A.,
  2004, \mn@doi [\apj] {10.1086/423444}, \href
  {http://adsabs.harvard.edu/abs/2004ApJ...614...17G} {614, 17}

\bibitem[\protect\citeauthoryear{{Gonzalez}, {Lacey}, {Baugh}  \&
  {Frenk}}{{Gonzalez} et~al.}{2011}]{Gonzalez2011}
{Gonzalez} J.~E.,  {Lacey} C.~G.,  {Baugh} C.~M.,   {Frenk} C.~S.,  2011,
  \mn@doi [\mnras] {10.1111/j.1365-2966.2010.18169.x}, \href
  {http://adsabs.harvard.edu/abs/2011MNRAS.413..749G} {413, 749}

\bibitem[\protect\citeauthoryear{{Granato}, {De Zotti}, {Silva}, {Bressan}  \&
  {Danese}}{{Granato} et~al.}{2004}]{Granato2004}
{Granato} G.~L.,  {De Zotti} G.,  {Silva} L.,  {Bressan} A.,   {Danese} L.,
  2004, \mn@doi [\apj] {10.1086/379875}, \href
  {http://adsabs.harvard.edu/abs/2004ApJ...600..580G} {600, 580}

\bibitem[\protect\citeauthoryear{{Granato}, {Silva}, {Lapi}, {Shankar}, {De
  Zotti}  \& {Danese}}{{Granato} et~al.}{2006}]{Granato2006}
{Granato} G.~L.,  {Silva} L.,  {Lapi} A.,  {Shankar} F.,  {De Zotti} G.,
  {Danese} L.,  2006, \mn@doi [\mnras] {10.1111/j.1745-3933.2006.00160.x},
  \href {http://adsabs.harvard.edu/abs/2006MNRAS.368L..72G} {368, L72}

\bibitem[\protect\citeauthoryear{{Grogin} et~al.,}{{Grogin}
  et~al.}{2011}]{Grogin2011}
{Grogin} N.~A.,  et~al., 2011, \mn@doi [\apjs] {10.1088/0067-0049/197/2/35},
  \href {http://adsabs.harvard.edu/abs/2011ApJS..197...35G} {197, 35}

\bibitem[\protect\citeauthoryear{{Gu}, {Conroy}  \& {Behroozi}}{{Gu}
  et~al.}{2016}]{Gu2016}
{Gu} M.,  {Conroy} C.,   {Behroozi} P.,  2016, preprint, \href
  {http://adsabs.harvard.edu/abs/2016arXiv160201099G} {} (\mn@eprint {arXiv}
  {1602.01099})

\bibitem[\protect\citeauthoryear{{Guo} et~al.,}{{Guo} et~al.}{2011}]{Guo2011}
{Guo} Q.,  et~al., 2011, \mn@doi [\mnras] {10.1111/j.1365-2966.2010.18114.x},
  \href {http://adsabs.harvard.edu/abs/2011MNRAS.413..101G} {413, 101}

\bibitem[\protect\citeauthoryear{{Hirschmann}, {Naab}, {Somerville}, {Burkert}
  \& {Oser}}{{Hirschmann} et~al.}{2012}]{Hirschmann2012}
{Hirschmann} M.,  {Naab} T.,  {Somerville} R.~S.,  {Burkert} A.,   {Oser} L.,
  2012, \mn@doi [\mnras] {10.1111/j.1365-2966.2011.19961.x}, \href
  {http://adsabs.harvard.edu/abs/2012MNRAS.419.3200H} {419, 3200}

\bibitem[\protect\citeauthoryear{{Hopkins} et~al.,}{{Hopkins}
  et~al.}{2010}]{Hopkins2010}
{Hopkins} P.~F.,  et~al., 2010, \mn@doi [\apj] {10.1088/0004-637X/724/2/915},
  \href {http://adsabs.harvard.edu/abs/2010ApJ...724..915H} {724, 915}

\bibitem[\protect\citeauthoryear{{Huertas-Company} et~al.,}{{Huertas-Company}
  et~al.}{2013}]{Huertas-Company2013}
{Huertas-Company} M.,  et~al., 2013, \mn@doi [\mnras] {10.1093/mnras/sts150},
  \href {http://adsabs.harvard.edu/abs/2013MNRAS.428.1715H} {428, 1715}

\bibitem[\protect\citeauthoryear{{Huertas-Company} et~al.,}{{Huertas-Company}
  et~al.}{2015}]{Huertas-Company2015}
{Huertas-Company} M.,  et~al., 2015, \mn@doi [\apj]
  {10.1088/0004-637X/809/1/95}, \href
  {http://adsabs.harvard.edu/abs/2015ApJ...809...95H} {809, 95}

\bibitem[\protect\citeauthoryear{{Jiang} \& {van den Bosch}}{{Jiang} \& {van
  den Bosch}}{2014}]{Jiang2014}
{Jiang} F.,  {van den Bosch} F.~C.,  2014, \mn@doi [\mnras]
  {10.1093/mnras/stu280}, \href
  {http://adsabs.harvard.edu/abs/2014MNRAS.440..193J} {440, 193}

\bibitem[\protect\citeauthoryear{{Klypin}, {Trujillo-Gomez}  \&
  {Primack}}{{Klypin} et~al.}{2011}]{Klypin2011}
{Klypin} A.~A.,  {Trujillo-Gomez} S.,   {Primack} J.,  2011, \mn@doi [\apj]
  {10.1088/0004-637X/740/2/102}, \href
  {http://adsabs.harvard.edu/abs/2011ApJ...740..102K} {740, 102}

\bibitem[\protect\citeauthoryear{{Klypin}, {Yepes}, {Gottl{\"o}ber}, {Prada}
  \& {He{\ss}}}{{Klypin} et~al.}{2016}]{Klypin2016}
{Klypin} A.,  {Yepes} G.,  {Gottl{\"o}ber} S.,  {Prada} F.,   {He{\ss}} S.,
  2016, \mn@doi [\mnras] {10.1093/mnras/stw248}, \href
  {http://adsabs.harvard.edu/abs/2016MNRAS.457.4340K} {457, 4340}

\bibitem[\protect\citeauthoryear{{Kravtsov}, {Vikhlinin}  \&
  {Meshscheryakov}}{{Kravtsov} et~al.}{2014}]{Kravtsov2014}
{Kravtsov} A.,  {Vikhlinin} A.,   {Meshscheryakov} A.,  2014, preprint, \href
  {http://adsabs.harvard.edu/abs/2014arXiv1401.7329K} {} (\mn@eprint {arXiv}
  {1401.7329})

\bibitem[\protect\citeauthoryear{{Lacey} et~al.,}{{Lacey}
  et~al.}{2015}]{Lacey2015}
{Lacey} C.~G.,  et~al., 2015, preprint, \href
  {http://adsabs.harvard.edu/abs/2015arXiv150908473L} {} (\mn@eprint {arXiv}
  {1509.08473})

\bibitem[\protect\citeauthoryear{{Lapi}, {Gonzalez Nuevo}  \& {Fan}}{{Lapi}
  et~al.}{2011}]{Lapi2011}
{Lapi} A.,  {Gonzalez Nuevo} J.,   {Fan} L.,  2011, \mn@doi [\apj]
  {10.1088/0004-637X/742/1/24}, \href
  {http://adsabs.harvard.edu/abs/2011ApJ...742...24L} {742, 24}

\bibitem[\protect\citeauthoryear{{Lapi}, {Salucci}  \& {Danese}}{{Lapi}
  et~al.}{2013}]{Lapi2013}
{Lapi} A.,  {Salucci} P.,   {Danese} L.,  2013, \mn@doi [\apj]
  {10.1088/0004-637X/772/2/85}, \href
  {http://adsabs.harvard.edu/abs/2013ApJ...772...85L} {772, 85}

\bibitem[\protect\citeauthoryear{{Laporte} \& {White}}{{Laporte} \&
  {White}}{2015}]{Laporte2015}
{Laporte} C.~F.~P.,  {White} S.~D.~M.,  2015, \mn@doi [\mnras]
  {10.1093/mnras/stv112}, \href
  {http://adsabs.harvard.edu/abs/2015MNRAS.451.1177L} {451, 1177}

\bibitem[\protect\citeauthoryear{{Leauthaud} et~al.,}{{Leauthaud}
  et~al.}{2016}]{Leauthaud2016}
{Leauthaud} A.,  et~al., 2016, \mn@doi [\mnras] {10.1093/mnras/stw117}, \href
  {http://adsabs.harvard.edu/abs/2016MNRAS.457.4021L} {457, 4021}

\bibitem[\protect\citeauthoryear{{Lidman} et~al.,}{{Lidman}
  et~al.}{2012}]{Lidman2012}
{Lidman} C.,  et~al., 2012, \mn@doi [\mnras]
  {10.1111/j.1365-2966.2012.21984.x}, \href
  {http://adsabs.harvard.edu/abs/2012MNRAS.427..550L} {427, 550}

\bibitem[\protect\citeauthoryear{{Maraston} et~al.,}{{Maraston}
  et~al.}{2013}]{Maraston2013}
{Maraston} C.,  et~al., 2013, \mn@doi [\mnras] {10.1093/mnras/stt1424}, \href
  {http://adsabs.harvard.edu/abs/2013MNRAS.435.2764M} {435, 2764}

\bibitem[\protect\citeauthoryear{{Marchesini}, {van Dokkum}, {F{\"o}rster
  Schreiber}, {Franx}, {Labb{\'e}}  \& {Wuyts}}{{Marchesini}
  et~al.}{2009}]{Marchesini2009}
{Marchesini} D.,  {van Dokkum} P.~G.,  {F{\"o}rster Schreiber} N.~M.,  {Franx}
  M.,  {Labb{\'e}} I.,   {Wuyts} S.,  2009, \mn@doi [\apj]
  {10.1088/0004-637X/701/2/1765}, \href
  {http://adsabs.harvard.edu/abs/2009ApJ...701.1765M} {701, 1765}

\bibitem[\protect\citeauthoryear{{Marchesini} et~al.,}{{Marchesini}
  et~al.}{2014}]{Marchesini2014}
{Marchesini} D.,  et~al., 2014, \mn@doi [\apj] {10.1088/0004-637X/794/1/65},
  \href {http://adsabs.harvard.edu/abs/2014ApJ...794...65M} {794, 65}

\bibitem[\protect\citeauthoryear{{McDermid}, {Alatalo}  \& {Blitz}}{{McDermid}
  et~al.}{2015}]{McDermid2015}
{McDermid} R.~M.,  {Alatalo} K.,   {Blitz} L.,  2015, \mn@doi [\mnras]
  {10.1093/mnras/stv105}, \href
  {http://adsabs.harvard.edu/abs/2015MNRAS.448.3484M} {448, 3484}

\bibitem[\protect\citeauthoryear{{Mendel} et~al.,}{{Mendel}
  et~al.}{2015}]{Mendel2015}
{Mendel} J.~T.,  et~al., 2015, \mn@doi [\apjl] {10.1088/2041-8205/804/1/L4},
  \href {http://adsabs.harvard.edu/abs/2015ApJ...804L...4M} {804, L4}

\bibitem[\protect\citeauthoryear{{Merlin}, {Chiosi}, {Piovan}, {Grassi},
  {Buonomo}  \& {Barbera}}{{Merlin} et~al.}{2012}]{Merlin2012}
{Merlin} E.,  {Chiosi} C.,  {Piovan} L.,  {Grassi} T.,  {Buonomo} U.,
  {Barbera} F.~L.,  2012, \mn@doi [\mnras] {10.1111/j.1365-2966.2012.21965.x},
  \href {http://adsabs.harvard.edu/abs/2012MNRAS.427.1530M} {427, 1530}

\bibitem[\protect\citeauthoryear{{Mo}, {van den Bosch}  \& {White}}{{Mo}
  et~al.}{2010}]{Mo2010}
{Mo} H.,  {van den Bosch} F.~C.,   {White} S.,  2010, {Galaxy Formation and
  Evolution}

\bibitem[\protect\citeauthoryear{{Montes}, {Trujillo}, {Prieto}  \&
  {Acosta-Pulido}}{{Montes} et~al.}{2014}]{Montes2014}
{Montes} M.,  {Trujillo} I.,  {Prieto} M.~A.,   {Acosta-Pulido} J.~A.,  2014,
  \mn@doi [\mnras] {10.1093/mnras/stu037}, \href
  {http://adsabs.harvard.edu/abs/2014MNRAS.439..990M} {439, 990}

\bibitem[\protect\citeauthoryear{{Moster}, {Naab}  \& {White}}{{Moster}
  et~al.}{2013}]{Moster2013}
{Moster} B.~P.,  {Naab} T.,   {White} S.~D.~M.,  2013, \mn@doi [\mnras]
  {10.1093/mnras/sts261}, \href
  {http://adsabs.harvard.edu/abs/2013MNRAS.428.3121M} {428, 3121}

\bibitem[\protect\citeauthoryear{{Naab}, {Johansson}  \& {Ostriker}}{{Naab}
  et~al.}{2009}]{Naab2009}
{Naab} T.,  {Johansson} P.~H.,   {Ostriker} J.~P.,  2009, \mn@doi [\apjl]
  {10.1088/0004-637X/699/2/L178}, \href
  {http://adsabs.harvard.edu/abs/2009ApJ...699L.178N} {699, L178}

\bibitem[\protect\citeauthoryear{{Newman}, {Ellis}, {Andreon}, {Treu},
  {Raichoor}  \& {Trinchieri}}{{Newman} et~al.}{2014}]{Newman2014}
{Newman} A.~B.,  {Ellis} R.~S.,  {Andreon} S.,  {Treu} T.,  {Raichoor} A.,
  {Trinchieri} G.,  2014, \mn@doi [\apj] {10.1088/0004-637X/788/1/51}, \href
  {http://adsabs.harvard.edu/abs/2014ApJ...788...51N} {788, 51}

\bibitem[\protect\citeauthoryear{{Oliva-Altamirano}, {Brough}, {Jimmy},
  {Couch}, {McDermid}, {Lidman}, {von der Linden}  \&
  {Sharp}}{{Oliva-Altamirano} et~al.}{2015}]{Oliva-Altamirano2015}
{Oliva-Altamirano} P.,  {Brough} S.,  {Jimmy} Kim-Vy T.,  {Couch} W.~J.,
  {McDermid} R.~M.,  {Lidman} C.,  {von der Linden} A.,   {Sharp} R.,  2015,
  \mn@doi [\mnras] {10.1093/mnras/stv475}, \href
  {http://adsabs.harvard.edu/abs/2015MNRAS.449.3347O} {449, 3347}

\bibitem[\protect\citeauthoryear{{Oser}, {Ostriker}, {Naab}, {Johansson}  \&
  {Burkert}}{{Oser} et~al.}{2010}]{Oser2010}
{Oser} L.,  {Ostriker} J.~P.,  {Naab} T.,  {Johansson} P.~H.,   {Burkert} A.,
  2010, \mn@doi [\apj] {10.1088/0004-637X/725/2/2312}, \href
  {http://adsabs.harvard.edu/abs/2010ApJ...725.2312O} {725, 2312}

\bibitem[\protect\citeauthoryear{{Peeples} \& {Somerville}}{{Peeples} \&
  {Somerville}}{2013}]{Peeples2013}
{Peeples} M.~S.,  {Somerville} R.~S.,  2013, \mn@doi [\mnras]
  {10.1093/mnras/sts158}, \href
  {http://adsabs.harvard.edu/abs/2013MNRAS.428.1766P} {428, 1766}

\bibitem[\protect\citeauthoryear{{Peng}, {Lilly}  \& {Kova{\v c}}}{{Peng}
  et~al.}{2010}]{Peng2010}
{Peng} Y.-j.,  {Lilly} S.~J.,   {Kova{\v c}} K.,  2010, \mn@doi [\apj]
  {10.1088/0004-637X/721/1/193}, \href
  {http://adsabs.harvard.edu/abs/2010ApJ...721..193P} {721, 193}

\bibitem[\protect\citeauthoryear{{Pipino}, {Devriendt}, {Thomas}, {Silk}  \&
  {Kaviraj}}{{Pipino} et~al.}{2009}]{Pipino2009}
{Pipino} A.,  {Devriendt} J.~E.~G.,  {Thomas} D.,  {Silk} J.,   {Kaviraj} S.,
  2009, \mn@doi [\aap] {10.1051/0004-6361/200811269}, \href
  {http://adsabs.harvard.edu/abs/2009A%26A...505.1075P} {505, 1075}

\bibitem[\protect\citeauthoryear{{Poggianti} et~al.,}{{Poggianti}
  et~al.}{2006}]{Poggianti2006}
{Poggianti} B.~M.,  et~al., 2006, \mn@doi [\apj] {10.1086/500666}, \href
  {http://adsabs.harvard.edu/abs/2006ApJ...642..188P} {642, 188}

\bibitem[\protect\citeauthoryear{{Posti}, {Nipoti}, {Stiavelli}  \&
  {Ciotti}}{{Posti} et~al.}{2014}]{Posti2014}
{Posti} L.,  {Nipoti} C.,  {Stiavelli} M.,   {Ciotti} L.,  2014, \mn@doi
  [\mnras] {10.1093/mnras/stu301}, \href
  {http://adsabs.harvard.edu/abs/2014MNRAS.440..610P} {440, 610}

\bibitem[\protect\citeauthoryear{{Ragone-Figueroa} \&
  {Granato}}{{Ragone-Figueroa} \& {Granato}}{2011}]{Ragone-Figueroa2011}
{Ragone-Figueroa} C.,  {Granato} G.~L.,  2011, \mn@doi [\mnras]
  {10.1111/j.1365-2966.2011.18670.x}, \href
  {http://adsabs.harvard.edu/abs/2011MNRAS.414.3690R} {414, 3690}

\bibitem[\protect\citeauthoryear{{Saglia} et~al.,}{{Saglia}
  et~al.}{2010}]{Saglia2010}
{Saglia} R.~P.,  et~al., 2010, \mn@doi [\aap] {10.1051/0004-6361/201014703},
  \href {http://adsabs.harvard.edu/abs/2010A%26A...524A...6S} {524, A6}

\bibitem[\protect\citeauthoryear{{Scoville}, {Aussel}  \& {Brusa}}{{Scoville}
  et~al.}{2007}]{Scoville2007}
{Scoville} N.,  {Aussel} H.,   {Brusa} M.,  2007, \mn@doi [\apjs]
  {10.1086/516585}, \href {http://adsabs.harvard.edu/abs/2007ApJS..172....1S}
  {172, 1}

\bibitem[\protect\citeauthoryear{{Shankar} \& {Bernardi}}{{Shankar} \&
  {Bernardi}}{2009}]{Shankar2009}
{Shankar} F.,  {Bernardi} M.,  2009, \mn@doi [\mnras]
  {10.1111/j.1745-3933.2009.00665.x}, \href
  {http://adsabs.harvard.edu/abs/2009MNRAS.396L..76S} {396, L76}

\bibitem[\protect\citeauthoryear{{Shankar}, {Marulli}, {Bernardi}, {Mei},
  {Meert}  \& {Vikram}}{{Shankar} et~al.}{2013}]{Shankar2013}
{Shankar} F.,  {Marulli} F.,  {Bernardi} M.,  {Mei} S.,  {Meert} A.,   {Vikram}
  V.,  2013, \mn@doi [\mnras] {10.1093/mnras/sts001}, \href
  {http://adsabs.harvard.edu/abs/2013MNRAS.428..109S} {428, 109}

\bibitem[\protect\citeauthoryear{{Shankar} et~al.,}{{Shankar}
  et~al.}{2014a}]{Shankar2014a}
{Shankar} F.,  et~al., 2014a, \mn@doi [\mnras] {10.1093/mnras/stt2470}, \href
  {http://adsabs.harvard.edu/abs/2014MNRAS.439.3189S} {439, 3189}

\bibitem[\protect\citeauthoryear{{Shankar}, {Guo}  \& {Bouillot}}{{Shankar}
  et~al.}{2014b}]{Shankar2014b}
{Shankar} F.,  {Guo} H.,   {Bouillot} 2014b, \mn@doi [\apjl]
  {10.1088/2041-8205/797/2/L27}, \href
  {http://adsabs.harvard.edu/abs/2014ApJ...797L..27S} {797, L27}

\bibitem[\protect\citeauthoryear{{Shankar} et~al.,}{{Shankar}
  et~al.}{2015}]{Shankar2015}
{Shankar} F.,  et~al., 2015, \mn@doi [\apj] {10.1088/0004-637X/802/2/73}, \href
  {http://adsabs.harvard.edu/abs/2015ApJ...802...73S} {802, 73}

\bibitem[\protect\citeauthoryear{{Shankar} et~al.,}{{Shankar}
  et~al.}{2016}]{Shankar2016}
{Shankar} F.,  et~al., 2016, \mn@doi [\mnras] {10.1093/mnras/stw678}, \href
  {http://adsabs.harvard.edu/abs/2016MNRAS.tmp..465S} {}

\bibitem[\protect\citeauthoryear{{Shen}, {Mo}, {White}, {Blanton}, {Kauffmann},
  {Voges}, {Brinkmann}  \& {Csabai}}{{Shen} et~al.}{2003}]{Shen2003}
{Shen} S.,  {Mo} H.~J.,  {White} S.~D.~M.,  {Blanton} M.~R.,  {Kauffmann} G.,
  {Voges} W.,  {Brinkmann} J.,   {Csabai} I.,  2003, \mn@doi [\mnras]
  {10.1046/j.1365-8711.2003.06740.x}, \href
  {http://adsabs.harvard.edu/abs/2003MNRAS.343..978S} {343, 978}

\bibitem[\protect\citeauthoryear{{Smail}, {Chapman}, {Ivison}, {Blain},
  {Takata}, {Heckman}, {Dunlop}  \& {Sekiguchi}}{{Smail}
  et~al.}{2003}]{Smail2003}
{Smail} I.,  {Chapman} S.~C.,  {Ivison} R.~J.,  {Blain} A.~W.,  {Takata} T.,
  {Heckman} T.~M.,  {Dunlop} J.~S.,   {Sekiguchi} K.,  2003, \mn@doi [\mnras]
  {10.1046/j.1365-8711.2003.06621.x}, \href
  {http://adsabs.harvard.edu/abs/2003MNRAS.342.1185S} {342, 1185}

\bibitem[\protect\citeauthoryear{{Springel}}{{Springel}}{2005}]{Springel2005}
{Springel} V.,  2005, \mn@doi [\mnras] {10.1111/j.1365-2966.2005.09655.x},
  \href {http://adsabs.harvard.edu/abs/2005MNRAS.364.1105S} {364, 1105}

\bibitem[\protect\citeauthoryear{{Stewart}, {Bullock}, {Wechsler}  \&
  {Maller}}{{Stewart} et~al.}{2009}]{Stewart2009}
{Stewart} K.~R.,  {Bullock} J.~S.,  {Wechsler} R.~H.,   {Maller} A.~H.,  2009,
  \mn@doi [\apj] {10.1088/0004-637X/702/1/307}, \href
  {http://adsabs.harvard.edu/abs/2009ApJ...702..307S} {702, 307}

\bibitem[\protect\citeauthoryear{{Strazzullo} et~al.,}{{Strazzullo}
  et~al.}{2013}]{Strazzullo2013}
{Strazzullo} V.,  et~al., 2013, \mn@doi [\apj] {10.1088/0004-637X/772/2/118},
  \href {http://adsabs.harvard.edu/abs/2013ApJ...772..118S} {772, 118}

\bibitem[\protect\citeauthoryear{{Stringer}, {Trujillo}, {Dalla Vecchia}  \&
  {Martinez-Valpuesta}}{{Stringer} et~al.}{2015}]{Stringer2015}
{Stringer} M.,  {Trujillo} I.,  {Dalla Vecchia} C.,   {Martinez-Valpuesta} I.,
  2015, \mn@doi [\mnras] {10.1093/mnras/stv455}, \href
  {http://adsabs.harvard.edu/abs/2015MNRAS.449.2396S} {449, 2396}

\bibitem[\protect\citeauthoryear{{Swinbank} et~al.,}{{Swinbank}
  et~al.}{2005}]{Swinbank2005}
{Swinbank} A.~M.,  et~al., 2005, \mn@doi [\mnras]
  {10.1111/j.1365-2966.2005.08901.x}, \href
  {http://adsabs.harvard.edu/abs/2005MNRAS.359..401S} {359, 401}

\bibitem[\protect\citeauthoryear{{Thanjavur}, {Simard}, {Bluck}  \&
  {Mendel}}{{Thanjavur} et~al.}{2016}]{Thanjavur2016}
{Thanjavur} K.,  {Simard} L.,  {Bluck} A.~F.~L.,   {Mendel} T.,  2016, \mn@doi
  [\mnras] {10.1093/mnras/stw495}, \href
  {http://adsabs.harvard.edu/abs/2016MNRAS.459...44T} {459, 44}

\bibitem[\protect\citeauthoryear{{Thomas}, {Maraston}, {Bender}  \& {Mendes de
  Oliveira}}{{Thomas} et~al.}{2005}]{Thomas2005}
{Thomas} D.,  {Maraston} C.,  {Bender} R.,   {Mendes de Oliveira} C.,  2005,
  \mn@doi [\apj] {10.1086/426932}, \href
  {http://adsabs.harvard.edu/abs/2005ApJ...621..673T} {621, 673}

\bibitem[\protect\citeauthoryear{{Tomczak} et~al.,}{{Tomczak}
  et~al.}{2016}]{Tomczak2016}
{Tomczak} A.~R.,  et~al., 2016, \mn@doi [\apj] {10.3847/0004-637X/817/2/118},
  \href {http://adsabs.harvard.edu/abs/2016ApJ...817..118T} {817, 118}

\bibitem[\protect\citeauthoryear{{Torrey} et~al.,}{{Torrey}
  et~al.}{2015}]{Torrey2015}
{Torrey} P.,  et~al., 2015, \mn@doi [\mnras] {10.1093/mnras/stv1986}, \href
  {http://adsabs.harvard.edu/abs/2015MNRAS.454.2770T} {454, 2770}

\bibitem[\protect\citeauthoryear{{Vincoletto}, {Matteucci}, {Calura}, {Silva}
  \& {Granato}}{{Vincoletto} et~al.}{2012}]{Vincoletto2012}
{Vincoletto} L.,  {Matteucci} F.,  {Calura} F.,  {Silva} L.,   {Granato} G.,
  2012, \mn@doi [\mnras] {10.1111/j.1365-2966.2012.20535.x}, \href
  {http://adsabs.harvard.edu/abs/2012MNRAS.421.3116V} {421, 3116}

\bibitem[\protect\citeauthoryear{{Welker}, {Dubois}, {Devriendt}, {Pichon},
  {Kaviraj}  \& {Peirani}}{{Welker} et~al.}{2015}]{Welker2015}
{Welker} C.,  {Dubois} Y.,  {Devriendt} J.,  {Pichon} C.,  {Kaviraj} S.,
  {Peirani} S.,  2015, preprint, \href
  {http://adsabs.harvard.edu/abs/2015arXiv150205053W} {} (\mn@eprint {arXiv}
  {1502.05053})

\bibitem[\protect\citeauthoryear{{White} \& {Frenk}}{{White} \&
  {Frenk}}{1991}]{White1991}
{White} S.~D.~M.,  {Frenk} C.~S.,  1991, \mn@doi [\apj] {10.1086/170483}, \href
  {http://adsabs.harvard.edu/abs/1991ApJ...379...52W} {379, 52}

\bibitem[\protect\citeauthoryear{{Wilman}, {Fontanot}, {De Lucia}, {Erwin}  \&
  {Monaco}}{{Wilman} et~al.}{2013}]{Wilman2013}
{Wilman} D.~J.,  {Fontanot} F.,  {De Lucia} G.,  {Erwin} P.,   {Monaco} P.,
  2013, \mn@doi [\mnras] {10.1093/mnras/stt941}, \href
  {http://adsabs.harvard.edu/abs/2013MNRAS.433.2986W} {433, 2986}

\bibitem[\protect\citeauthoryear{{Woo}, {Dekel}, {Faber}  \& {Koo}}{{Woo}
  et~al.}{2015}]{Woo2015}
{Woo} J.,  {Dekel} A.,  {Faber} S.~M.,   {Koo} D.~C.,  2015, \mn@doi [\mnras]
  {10.1093/mnras/stu2755}, \href
  {http://adsabs.harvard.edu/abs/2015MNRAS.448..237W} {448, 237}

\bibitem[\protect\citeauthoryear{{Zolotov} et~al.,}{{Zolotov}
  et~al.}{2015}]{Zolotov2015}
{Zolotov} A.,  et~al., 2015, \mn@doi [\mnras] {10.1093/mnras/stv740}, \href
  {http://adsabs.harvard.edu/abs/2015MNRAS.450.2327Z} {450, 2327}

\bibitem[\protect\citeauthoryear{{van Dokkum} \& {Franx}}{{van Dokkum} \&
  {Franx}}{1996}]{VanDokkum1996}
{van Dokkum} P.~G.,  {Franx} M.,  1996, \mnras, \href
  {http://adsabs.harvard.edu/abs/1996MNRAS.281..985V} {281, 985}

\bibitem[\protect\citeauthoryear{{van Dokkum} et~al.,}{{van Dokkum}
  et~al.}{2010}]{VanDokkum2010}
{van Dokkum} P.~G.,  et~al., 2010, \mn@doi [\apj]
  {10.1088/0004-637X/709/2/1018}, \href
  {http://adsabs.harvard.edu/abs/2010ApJ...709.1018V} {709, 1018}

\bibitem[\protect\citeauthoryear{{van der Wel}, {Franx}  \& {van Dokkum}}{{van
  der Wel} et~al.}{2014}]{VanderWel2014}
{van der Wel} A.,  {Franx} M.,   {van Dokkum} 2014, \mn@doi [\apj]
  {10.1088/0004-637X/788/1/28}, \href
  {http://adsabs.harvard.edu/abs/2014ApJ...788...28V} {788, 28}

\makeatother
\end{thebibliography}

\bsp

\label{lastpage}

\end{document}